\documentclass[prd,reprint,onecolumn,superscriptaddress,amsfonts,nofootinbib]{revtex4-2}
\usepackage{amsmath,amssymb}

\usepackage{slashed}
\usepackage{graphicx}
\usepackage{epsfig}
\usepackage{epstopdf}
\usepackage{adjustbox}

\newcommand{\ii}{\mathrm{i}}

\newcommand{\x}{\mathsf{x}}
\newcommand{\te}{\mathsf{te}}
\newcommand{\tmm}{\mathsf{tm}}

\usepackage{xcolor} \definecolor{darkgreen}{rgb}{0,.5,0}
\usepackage[colorlinks,filecolor=blue,citecolor=darkgreen,unicode]{hyperref}

\begin{document}
\title{
Bulk contributions to the Casimir interaction of Dirac materials}
\author{M. Bordag}\email{bordag@uni-leipzig.de}
\affiliation{Leipzig University, Institute for Theoretical Physics,  04109 Leipzig, Germany}
\author{I. Fialkovsky}
\affiliation{CMCC-Universidade Federal do ABC, Santo Andr\'e, S.P., Brazil}
\author{N. Khusnutdinov}\email{nail.khusnutdinov@gmail.com}
\affiliation{CMCC-Universidade Federal do ABC, Santo Andr\'e, S.P., Brazil}
\author{D. Vassilevich}\email{dvassil@gmail.com}
\affiliation{CMCC-Universidade Federal do ABC, Santo Andr\'e, S.P., Brazil}
\affiliation{Physics Department, Tomsk State University, Tomsk, Russia}

\begin{abstract}
Exploiting methods of Quantum Field Theory we compute the bulk polarization tensor and bulk dielectric functions for Dirac materials in the presence of a mass gap, chemical potential, and finite temperature. Using these results (and neglecting eventual boundary effects), we study the Casimir interaction of Dirac materials. We describe in detail the characteristic features of the dielectric functions and their influence on the Casimir pressure.
\end{abstract}
\maketitle

\section{Introduction}\label{sec:Intro}
Topological insulators are an actual  and, by now, a quite well known topic, see \cite{Qi:2011zya,Hasan:2010xy}. The characteristic feature of these materials is the presence of topologically protected surface states. (We shall consider here  3D materials only). The simplest continuous model having this property is a massive Dirac fermion subjected to suitable boundary conditions. 

Over the recent years, the physics of Casimir interaction \cite{Bordag:2009zzd} emerged as an important instrument for the study of properties of new materials \cite{Woods:2015pla,Lu:2021jvu}. By this motivation, in the present work we investigate the Casimir interaction between Dirac materials. Thereby we restrict ourselves to the account of the bulk properties of these materials on the Casimir force as a first step and intend to account for the boundary effects later. 
The publications on Casimir interactions of various Dirac materials have been reviewed in \cite{Woods:2015pla,Lu:2021jvu}. We like to mention a couple of recent papers, \cite{Rodriguez-Lopez:2020:scorCitIaIIWs,Farias:2020qqp,Ishikawa:2020icy}. However, the papers which have been published so far use different physical setups making any direct comparison to the results of the present paper practically  impossible.

The bulk electronic properties will be expressed in terms of the polarization tensor of Dirac fermions and computed using te methods of Quantum Field Theory (QFT) as one-loop Feynman diagram. 
This approach was applied to a 2D Dirac material -- graphene -- in the papers \cite{Bordag:2009fz,Fialkovsky:2011pu} and subsequently confirmed at the experimental studies \cite{Banishev:2013,Klimchitskaya:2014axa,Liu:2021ice}. The polarization tensor approach is known  to take properly into account the spatial dispersion effects by construction. As has been demonstrated recently \cite{Klimchitskaya:2020qmy,Klimchitskaya:2021qri}, these effects are essential for a resolution of some internal controversies in the Casimir physics. We need the full expression for the polarization tensor at arbitrary values of four-momenta, mass, temperature, and chemical potential. Despite very large literature of finite-temperature QFT we were not able to find a suitable expression to required generality and had to redo this (rather standard) calculation. The dielectric functions $\varepsilon^l$ (longitudinal)  and $\varepsilon^t$ (transversal) are expressed through components of the polarization tensor with the help of Lindhard formalism \cite{Lindhard:1954}.
We mention closely related calculations in \cite{bord18-10-74}. There the Casimir interaction between half-spaces with a scalar field confined to them, interacting through another scalar field across the gap for scalar fields, was calculated using the 'TGTG'-formula. However, the methods for the calculation of the polarization tensor used there, are different from the ones used in the present paper.

Our purpose is not to study the Casimir interaction of a particular material, but rather to answer the question: What are the characteristic features of dielectric functions and Casimir pressure of a material having Dirac quasiparticles in the bulk? Therefore, we perform our study for a reasonably wide range of parameters. We find that quantum corrections to the magnetic permeability are always negligible. The dielectric functions $\varepsilon^{l,t}$ have a strong peak at low (imaginary) frequencies and momenta with rather characteristic dependence on $m$, $\mu$, and $T$, and a slow varying tail at moderate or high momenta and frequencies. The Casimir pressure between two identical Dirac materials is influenced significantly by quantum corrections. It is suppressed at short distances and low temperatures and enhanced at large separations and high temperatures. We trace this behaviour back to the properties of dielectric functions. We also find that almost the whole dependence of the Casimir pressure on $T$ and $\mu$ is due to the zero Matsubara frequency contribution to the Lifshitz sum.

This paper is organized as follows. The polarization tensor is computed in the next section while some details of this computation are placed in Appendix \ref{sec:Com}. In section \ref{sec:beh} we analyze the dielectric functions at imaginary frequencies. Section \ref{sec:Cas} is dedicated to the Casimir interaction. Some concluding remarks are contained in section \ref{sec:con}.

Throughout the paper we use natural units  $\hbar=c = k_B = 1$, if not stated otherwise. 

\section{Polarization tensor and form factors}\label{sec:Pol}
We consider an idealized material with quasiparticle excitations corresponding to one generation of Dirac fermions. Their spectrum is described by the Dirac operator
\begin{equation}
\slashed{D}=\ii \tilde{\gamma}^\nu (\partial_\nu +\ii e A_\nu) +m. \label{Dirop}
\end{equation}
Here $\nu=0,1,2,3$ is a four-vector index, $A_\nu$ is an electromagnetic potential.
A tilde over a four-vector means that the components are rescaled with the Fermi velocity $v_F$ as
\begin{equation}
\tilde\gamma^\mu\equiv \eta^{\mu}_\nu \gamma^\nu,\qquad \eta=\mathrm{diag}(1,v_F,v_F,v_F).
\end{equation}
The matrices $\gamma^\mu$ satisfy the usual Clifford relation $\gamma^\mu\gamma^\nu+\gamma^\nu\gamma^\mu =2g^{\mu\nu}$ with $g=\mathrm{diag}(+1,-1,-1,-1)$. Particular representation of the $\gamma$ matrices will play no role in what follows. Without any loss of generality we assume $m\geq 0$.

The interaction with quantum fermions leads to the following effective action in quadratic order for the electromagnetic potential
\begin{equation}
S_{\mathrm{eff}}=\frac 12 \int d^4x\, d^4y\, A_\mu(x) \Pi^{\mu\nu}(x,y)\, A_\nu(y),\label{Seff}
\end{equation}
where $\Pi^{\mu\nu}$ is the polarization tensor. Due to the translation invariance, it is convenient to make a Fourier transformation in (\ref{Seff}). Our conventions are clear from the formula
\begin{equation*}
A_\mu(x)=\int\frac{d^4k}{(2\pi)^4} e^{\ii kx}A_\mu(k).
\end{equation*}
Thus, we have
\begin{equation}
S_{\mathrm{eff}}=\frac 12 \int\frac{d^4k}{(2\pi)^4}A_\mu(-k) \Pi^{\mu\nu}(k)\, A_\nu(k).\label{Seffk}
\end{equation}

For zero temperature and zero chemical potential, $T=\mu=0$, and to the lowest order of perturbation expansion (one-loop) the polarization tensor, which will be denoted by $\Pi_0(k)$, can be written as
\begin{equation}
\Pi^{\mu\nu}_{0}(p)=\ii e^2\int \frac{d^4k}{(2\pi)^4} \mathrm{tr}\, \left[ \tilde\gamma^\mu \slashed{D}_0^{-1}(k)\gamma^\nu \slashed{D}_0^{-1}(k-p) \right].\label{Pmn1}
\end{equation}
Here $\slashed{D}_0^{-1}(k)$ is a Fourier transform of the inverse of free Dirac operator ($A_\mu=0$), 
\begin{equation}
\slashed{D}_0^{-1}(k)=-\frac{k_\mu\tilde\gamma^\mu +m}{k^2-m^2}.\label{DirInv}
\end{equation}
We use causal Greens' functions, so that the integration contour for $k_0$ is defined by the shift $k_0\to k_0+\ii 0\, \mathrm{sgn}\, (k_0)$.

To include a non-zero temperature $T$ and a non-zero chemical potential we shall use the imaginary time Matsubara formalism. First one has to shift the temporal components of the momenta of fermions by the chemical potential $\mu$, $k_0\to k_0+\mu$, without shifting the contour. Then one has to pass to imaginary Matsubara frequencies, $k_0\to \ii k_4$, $p_0\to\ii p_4$, $k_4=2\pi T(n+\tfrac 12)$, $p_4=2\pi T l$, $n,l\in \mathbb{Z}$. (Later we shall use a special notation $\xi_l=2\pi l T$ for the bosonic Matsubara frequencies). At the last step, one has to replace the integral over $k_0$ by a sum over the Matsubara frequencies,
\begin{equation}
\int_{-\infty}^{\infty}dk_0 \to 2\pi \ii  T\sum_{n\in\mathbb{Z}}.\label{InttoSum}
\end{equation}
Thus we obtain
\begin{equation}
\Pi^{\mu\nu}(p)=-e^2T\sum_{n\in\mathbb{Z}} \int \frac{d^3\vec{k}}{(2\pi)^3}
\mathrm{tr}\, \left[ \tilde{\gamma}^\mu S(k)\tilde{\gamma}^\nu S(k-p) \right],\label{PiEu}
\end{equation}
where all momenta are Euclidean as described above. $S(k)$ is the Euclidean propagator,
\begin{equation}
S(k)=-\frac{(\ii k_4+\mu)\gamma^0 +v_F\vec{k}\cdot\vec{\gamma}+m}{(\ii k_4+\mu)^2-v_F^2\vec{k}^2 -m^2}.\label{Sk}
\end{equation}
Our notations respect the natural position of indices, $\vec{k}=(k_1,k_2,k_3)$ while $\vec{\gamma}=(\gamma^1,\gamma^2,\gamma^3)$.
We would like to stress that we do not change the $\gamma$-matrices and do not make Euclidean rotation of the components of $\Pi^{\mu\nu}$.

As known, see for example \cite{Fialkovsky:2019rum}, the dependence of $\Pi$ on the Fermi velocity can be accounted for in a very simple way. One can make the change of integration variable $\vec{k}\to v_F\vec{k}$ in (\ref{PiEu}) to see that
\begin{equation}
\Pi^{\mu\nu}(p)=v_F^{-3}\eta^\mu_\rho \eta^\nu_\sigma \widehat{\Pi}^{\rho\sigma}(\tilde{p}),\label{vFPi}
\end{equation}
where $\widehat{\Pi}^{\mu\nu}(p)$ is the polarization tensor computed for $v_F=1$ and $\tilde{p}_\mu = \eta_\mu^\nu p_\nu$. As we shall see below, one needs just two specific combinations of the  components of $\widehat{\Pi}^{\mu\nu}$, namely $\widehat{\Pi}^{00}$ and $\widehat{\Pi}^{\mathrm{tr}}\equiv \widehat{\Pi}^\mu_\mu$. We introduce a shorthand notation $\widehat{\Pi}^{\x}$ for both of them  with either $\x=00$ or $\x=\mathrm{tr}$.

The computations of polarization tensor can be found in Appendix \ref{sec:Com} while the final result reads
\begin{equation}
		\widehat\Pi^{\x} (p) = 	\widehat\Pi^{\x}_0 (p) + 	\widehat\Pi^{\x}_{T,\mu} (p). \label{PiPP}
\end{equation}
The individual terms on the right hand side are given by
\begin{eqnarray}
	\widehat\Pi^{\x}_{T,\mu} (p) &=& +\frac{e^2}{2\pi^2} \int_0^\infty \frac{\vec{k}^2d|\vec{k}|}{E_k } \left\{I^\x_{+} + I^\x_{-}\right\} \Xi (E_k,\mu),\label{eq:PiTmu}\\
	\widehat\Pi^{\x}_{0} (p) &=& -\frac{e^2}{2\pi^2} \int_0^\infty \frac{\vec{k}^2d|\vec{k}|}{E_k } \left\{I^\x_{+} + I^\x_{-}\right\}.\label{eq:Pi0fin}
\end{eqnarray}
We used the following notations
\begin{equation}
		\Xi (E_k,\mu) = \frac{1}{e^{\frac{E_k +\mu}{T}}+1} + \frac{1}{e^{\frac{E_k -\mu}{T}}+1},\qquad E_k = \sqrt{m^2 + \vec{k}^2}, 
\end{equation}
and
\begin{eqnarray}
	I^{00}_{\pm} &=& 1 + \frac{4E_k^2 - \vec{p}^{\,2} - p_4^2 \pm  4\ii p_4 E_k}{4|\vec{k}||\vec{p}|}\ln \left[\frac{\vec{p}^{\,2} + p_4^2 + 2 |\vec{k}||\vec{p}| \mp 2 \ii p_4  E_k }{\vec{p}^{\,2} + p_4 ^2 - 2 |\vec{k}||\vec{p}| \mp 2 \ii p_4  E_k  }\right],\nonumber \\ 
	I^{\mathrm{tr}}_{\pm} &=& 2\left\{1 + \frac{2m^2  - \vec{p}^{\,2} - p_4 ^2}{4 |\vec{k}||\vec{p}|  }\ln \left[\frac{\vec{p}^{\,2} + p_4^2 + 2 |\vec{k}||\vec{p}| \mp 2 \ii p_4  E_k }{\vec{p}^{\,2} + p_4 ^2 - 2 |\vec{k}||\vec{p}| \mp 2 \ii p_4  E_k  }\right] \right\} .\label{eq:ISummary}
\end{eqnarray}
A characteristic property of the split (\ref{PiPP}) is that the tensor $\widehat{\Pi}_{0}$ depends neither on $T$ nor on $\mu$, while the part $\widehat{\Pi}_{T,\mu}$ vanishes when $T=\mu=0$. This motivates our choice of notations. The representation (\ref{PiPP}) has many other advantages which will be explained below. 

Let us now show how the full expression for polarization tensor can be recovered from  $\widehat{\Pi}^{\mathrm{tr}}$ and $\widehat{\Pi}^{00}$.
The polarization tensor for the problem in question has to satisfy a number of symmetry requirements. It has to be symmetric, invariant under spatial rotations and transversal. Thus, just two independent tensor structures are allowed:
\begin{eqnarray}
&&\Pi^{\mu\nu}(p)=\varphi_L(p)P_L^{\mu\nu}+\varphi_T(p)P_T^{\mu\nu},\label{PLPT} \\
&&P_L^{\mu\nu}=\frac{p^\mu p^\nu}{p^2} -\frac{p^\mu u^\nu +p^\nu u^\mu}{p_0} +\frac{u^\mu u^\nu p^2}{p_0^2},\qquad
P_T^{\mu\nu} =g^{\mu\nu} - \frac{p^\mu p^\nu}{p^2},\nonumber
\end{eqnarray}
where $u=(1,0,0,0)$ in the medium rest reference frame. The scalar functions $\varphi_L$ and $\varphi_T$ (form factors) can be expressed through the components of polarization tensor as
\begin{equation}
\varphi_T=\frac 12 \left( \frac{p^2}{\vec{p}^{\,2}}\, \Pi^{00} +\Pi_\mu^\mu \right),\qquad
\varphi_L=\frac {p_0^2}{2\vec{p}^{\,2}} \left( \frac{3p^2}{\vec{p}^{\,2}}\, \Pi^{00} +\Pi_\mu^\mu \right).
\label{vTvL}
\end{equation}
By using (\ref{vFPi}) we can express these quantities through the polarization tensor at $v_F=1$,
\begin{equation}
\Pi^{00}(p)=v_F^{-3}\widehat{\Pi}^{00}(\tilde p),\qquad \Pi_\mu^\mu(p) =v_F^{-3}( 1-v_F^2)\widehat{\Pi}^{00}(\tilde p)+v_F^{-1}\widehat{\Pi}_\mu^\mu(\tilde p).
\label{Pi-hatPi}
\end{equation}
Eqs. (\ref{vTvL})  and (\ref{Pi-hatPi}) are valid for both terms of the split (\ref{PiPP}) $\Pi_0$ and $\Pi_{T,\mu}$ individually. 

The integral over $|\vec{k}|$ in (\ref{eq:Pi0fin}) is divergent. Thus the polarization tensor $\Pi^{\mu\nu}$ needs to be regularized and renormalized. Note that $\Pi_{T,\mu}$ is finite, so that the renormalization can be performed at a zero temperature and a zero chemical potential, as expected. In the presence of $v_F$, the renormalization was performed recently in \cite{Fialkovsky:2019rum} in the Pauli-Villars formalism which we follow in this work (see also  \cite{Isobe:2012vh,Roy:2015zna,Pozo:2018yzs,Chernodub:2019blw} for a renormalization group analysis of such theories). A short summary of the procedure used in \cite{Fialkovsky:2019rum} is as follows. One uses subtraction of the contributions of Pauli-Villars regulator fields to polarization tensor. The expressions obtained are still divergent in the limit of infinitely massive regulators. These divergences are removed by a redefinition of bare dielectric permittivity $\varepsilon_0$ and magnetic permeability ${\mu}_{M,0}$ in the Maxwell action in a media
\begin{equation}
S_M=\frac 12 \int \frac{d^4p}{(2\pi)^4} \left[ \varepsilon_0\vec{E}(-p)\cdot \vec{E}(p) - {\mu}_{M,0}^{-1}\vec{B}(-p)\cdot \vec{B}(p)\right].\label{SM}
\end{equation}
According to the general philosophy of renormalization, $\varepsilon_0$ and ${\mu}_{M,0}$ do not depend on the momentum $p$. These two constants cannot be predicted in the framework of QFT and have to be considered as an input. The \emph{renormalized} $\Pi_0$ reads
\begin{equation}
\Pi^{\mu\nu}_0(p)=-\frac{e^2}{2\pi^2v_F^{3}}\eta_\sigma^\mu\eta_\rho^\nu[\tilde{p}^\sigma\tilde{p}^\rho-{g}^{\sigma\rho}\tilde{p}^2]\, f(\tilde{p}^2/m^2),\qquad f(z)\equiv\int_0^1dx\, x(1-x)\, \ln [1-zx(1-x)].\label{Pi0ren}
\end{equation}
To avoid a notation clutter, we do not introduce any special symbol for the renormalized tensor. From now on, only renormalized quantities will be used. Since $f(0)=0$, $\varepsilon_0$ and ${\mu}_{M,0}$ can be interpreted as a dielectric permittivity and a magnetic permeability, respectively, measured at vanishing temperature, zero chemical potential and zero external momentum. The quantum effective action (\ref{Seffk}) obtained in this way vanishes in the limit of infinitely large mass gap, $m\to\infty$, which is very natural from the physical point of view.

We use now the analysis by Lindhard \cite{Lindhard:1954} to relate these quantities to the components of dielectric tensor $\varepsilon^l$ and $\varepsilon^t$ which are more common in the condensed matter context. The matrix-valued dielectric function $\varepsilon_{ij}$ which relates $\vec{E}$ to the electric displacement $\vec{D}$,
$D_i=\varepsilon_{ij}E_j$, can be written through two scalar functions as
\begin{equation}
\varepsilon_{ij}(p)=\left( \delta_{ij} - \frac{p_ip_j}{\vec{p}^{\,2}}\right)\, \varepsilon^t (p)+ \frac {p_ip_j}{\vec{p}^{\,2}}\, \varepsilon^l(p).\label{epsij}
\end{equation}
By comparing the standard Maxwell equations to the one obtained by varying $S_M+S_{\mathrm{eff}}$ one comes to the following relations 
\begin{eqnarray}
&&\varepsilon^l=\varepsilon_0+\frac 1{p^2} \left( \frac{\vec{p}^{\,2}}{p_0^2} \varphi_L-\varphi_T\right)=\varepsilon_0 +\frac{\Pi^{00}}{\vec{p}^{\,2}},\label{epsl}\\
&&\varepsilon^t=\varepsilon_0-\frac{\vec{p}^{\,2}}{p_0^2} \left( \frac 1{\mu_{M,0}} -1\right) -\frac{\varphi_T}{p_0^2}=\varepsilon_0-\frac{\vec{p}^{\,2}}{p_0^2} \left( \frac 1{\mu_{M,0}} -1\right) -\frac{1}{2p_0^2} \left( \frac{p^2}{\vec{p}^{\,2}}\, \Pi^{00} +\Pi_\mu^\mu \right).\label{epst}
\end{eqnarray}
These two functions will play the central role in our work. At real frequencies, they define the optical properties of a bulk of a Dirac material. At imaginary frequencies, these functions enter the Lifshitz formula for Casimir energy.

\section{Behaviour of the dielectric functions}\label{sec:beh}
From now on we work with Euclidean momenta only. Let us introduce the following notations for the Euclidean norm of four-vectors
\begin{equation}
|p|:=\sqrt{p_4^2+\vec{p}^{\,2}},\qquad |\tilde{p}|:=\sqrt{p_4^2+v_F^2\vec{p}^{\,2}}.
\end{equation}
Although for $T\neq 0$, within the Matsubara formalism, the momentum $p_4$ takes only discrete values, $p_4=2\pi T l$ (see  Eq. (\ref{InttoSum})), the above expressions are valid for any imaginary frequency, $\omega=\ii p_4$, and the analysis below will be done for continuous values of $p_4$ regardless of the temperature. 

For the purpose of numerical study, we will need some numerical values for the constants characterizing Dirac materials. As we have already mentioned above, we are not going to stick to any particular material. We shall rather make our choice within some reasonable range. We fix $v_F=(600)^{-1}$. The bare dielectric permittivity $\varepsilon_0$ will be allowed to vary between $2$ and $10$, the mass will be taken $0.01$ eV or $0.1$ eV, while the chemical potential will take values between $\mu=0$ and $\mu=0.2$ eV.

First, consider the case $T=0=\mu$. 
By using (\ref{epsl}), (\ref{epst}) and the explicit form of polarization tensor (\ref{Pi0ren}) we obtain
\begin{eqnarray}
&&\varepsilon^l(p)=\varepsilon_0 - \frac{e^2}{2\pi^2 v_F}\, f(-|\tilde{p}|^2/m^2),\label{epl0}\\
&&\varepsilon^t(p)=\varepsilon_0 +\frac{\vec{p}^{\,2}}{p_4^2} \left( \frac{1}{\mu_{M,0}} -1\right) -\frac{e^2}{2\pi^2} \left( \frac 1{v_F} + \frac{v_F\vec{p}^{\,2}}{p_4^2} \right)\, f(-|\tilde{p}|^2/m^2).\label{ept0}
\end{eqnarray}
At this point, we observe that in the terms representing quantum corrections in (\ref{epl0}) and (\ref{ept0}), the spatial momentum $\vec{p}$ is always multiplied with $v_F$. This makes the spatial dispersion practically irrelevant except for the zeroth Matsubara frequency $p_4=\xi_0=0$. Therefore, instead of $\varepsilon^l$ and $\varepsilon^t$ one can use the dielectric permittivity and magnetic permeability which are given by the formulas \cite{Lindhard:1954,Landau8}
\begin{equation}
\varepsilon( p_4)=\lim_{\vec{p}\to 0}\varepsilon^t(p)=\lim_{\vec{p}\to 0}\varepsilon^l(p),\qquad 1-\frac{1}{\mu_{M}}=-\lim_{\vec{p}\to 0} \frac{p_4^2}{\vec{p}^{\,2}}(\varepsilon^t-\varepsilon^l),\label{eeemu}
\end{equation}
where in all function the Wick rotation $p_0=\ii p_4$ is understood. By using (\ref{Pi0ren}) we obtain
\begin{equation}
\varepsilon(p_4)=\varepsilon_0-\frac{e^2}{2\pi^2 v_F}\, f(-p_4^2/m^2),\qquad
\mu_{M}^{-1}(p_4)=\mu_{M,0}^{-1}-\frac{e^2v_F}{2\pi^2}\, f(-p_4^2/m^2).\label{eemumu}
\end{equation}

As compared to quantum corrections $\Delta_0\varepsilon(p_4)\equiv \varepsilon(p_4)-\varepsilon_0$ the corrections to $\Delta_0\mu_{M}^{-1}(p_4)\equiv\mu_{M}^{-1}(p_4)-\mu_{M,0}^{-1}$ are suppressed with $v_F^2$ which is a very small quantity. Thus, there is no significant correction to $\mu_{M,0}^{-1}$ for $T=\mu=0$.

Note that $f(-p_4^2/m^2)$ is a monotonously increasing function of $p_4^2$. Asymptotically, it behaves as $p_4^2/(30\, m^2)$ for small $p_4^2$ and as $\tfrac 16 \ln (p_4^2/m^2)$ for large $p_4^2$. The log-term may look troubling since it grows indefinitely for very large values of $p_4$. However, such terms are usual in QFT. They signal the necessity of a resummation of perturbation series. Fortunately, for the applications considered in the present paper this effect is not significant due to the presence of  an exponential damping in the Lifshitz formula for Casimir pressure, see Eq.\ (\ref{Lif}) below.

For $\mu\neq 0 \neq T$ the dielectric functions are given by complicated expressions. These expressions are invariant with respect to the replacement $\mu\to -\mu$. From now on we take $\mu\geq 0$. Most remarkable is the behavior of dielectric permittivity at small (imaginary) frequencies:\footnote{We use a notation $\varepsilon=\varepsilon_0+\Delta_0\varepsilon+\Delta_{T,\mu}\varepsilon$ consistent with the expansion (\ref{PiPP}) of polarization tensor.}
\begin{equation}
\Delta_{T,\mu}\varepsilon(p_4)\simeq \frac{\Omega_\varepsilon^2}{p_4^2}.\label{Plep}
\end{equation}
This term reminds us of the plasma model of dielectric permittivity with the plasma frequency given by
\begin{equation}
	\Omega_{\varepsilon}^2 = \frac{e^2}{3 \pi^2 v_F}  \int_0^{\infty} \frac{\vec{k}^2 d|\vec{k}|}{E_k}  \frac{2 E_k^2 + m^2}{E_k^2} \Xi (E_k, \mu).\label{Plfreq}
	\end{equation}
This integral can be evaluated in two cases. In the first one, when $\mu<m$ and $T$ is much smaller than $m-\mu$, one has	
\begin{equation}
		\Omega_{\varepsilon}^2  \simeq \frac{e^2 }{v_F} \sqrt{\frac{m T^3}{2\pi^3}} \left(e^{-\frac{m-\mu}{T}} + e^{-\frac{m+\mu}{T}}\right). \label{Plmum}
\end{equation}
If $m<\mu$ and $T\ll \mu-m$,
\begin{equation}
	\Omega_{\varepsilon}^2 \simeq \frac{e^2 (\mu^2-m^2)^{3/2} }{3\pi^2 v_F \mu} \left(1 + T^2 \frac{\pi^2}{6} \frac{2\mu^4+2m^4 - \mu^2 m^2}{\mu^2 (\mu^2 -m^2)^2}\right).\label{Plmmu} 
\end{equation}
In deriving this formula the low-temperature expansion from \cite{Khusnutdinov:2018ley} is useful.

The correction $\Delta_{T,\mu}\mu_{M}^{-1}$ for small imaginary frequencies has a form similar to (\ref{Plep}). It is suppressed by a factor of the order $v_F^2$ as compared to  $\Delta_{T,\mu}\varepsilon$. As one can check numerically, the corrections to $\mu_{M}^{-1}$ always remain several orders of magnitude smaller than the corrections to $\varepsilon$. This is consistent with the phenomenological observation that Dirac materials are non-magnetic which also prompts us to use $\mu_{M,0}=1$ till the end of this paper.

The formula (\ref{Plep}) is not valid at the zeroth Matsubara frequency $p_4=0$. At this point, the spatial dispersion is not negligible. Thus the full functions $\varepsilon^l$ and $\varepsilon^t$ have to be considered. The function $\varepsilon^l(0,\vec{p})$ has a pole at $|\vec{p}|=0$,
\begin{equation}
\varepsilon^l(0,\vec{p})\simeq \frac{e^2}{\pi^2 v_F^3 \vec{p}^{\,2}} \int_0^\infty \frac{d|\vec{k}|}{E_k } \left\{k^2 + E_k^2 \right\} 	\Xi (E_k,\mu). \label{el0mat}
\end{equation}
If $T$ is much smaller than both $m$ and $|m-\mu|$, the expression above can be simplified as
\begin{eqnarray}
&\varepsilon^l(0,\vec{p})\simeq \frac{e^2}{v_F^3\vec{p}^{\,2}}\sqrt{\frac{mT}{2\pi}}\left(e^{-\frac{m-\mu}{T}} + e^{-\frac{m+\mu}{T}}\right)\qquad & \text{for}\ m>\mu, \label{elmum}\\
&\varepsilon^l(0,\vec{p})\simeq \frac{e^2\mu}{\pi v_F^2 \vec{p}^{\,2}}\sqrt{\mu^2-m^2}\quad &\text{for}\ m<\mu .\label{elmmu}
\end{eqnarray}
One sees many similarities to the behaviour of pole term in $\varepsilon$, see (\ref{Plep}), (\ref{Plfreq}), (\ref{Plmmu}), and (\ref{Plmum}) above.

\begin{figure}[h]
	\centering
	\includegraphics[width=5cm]{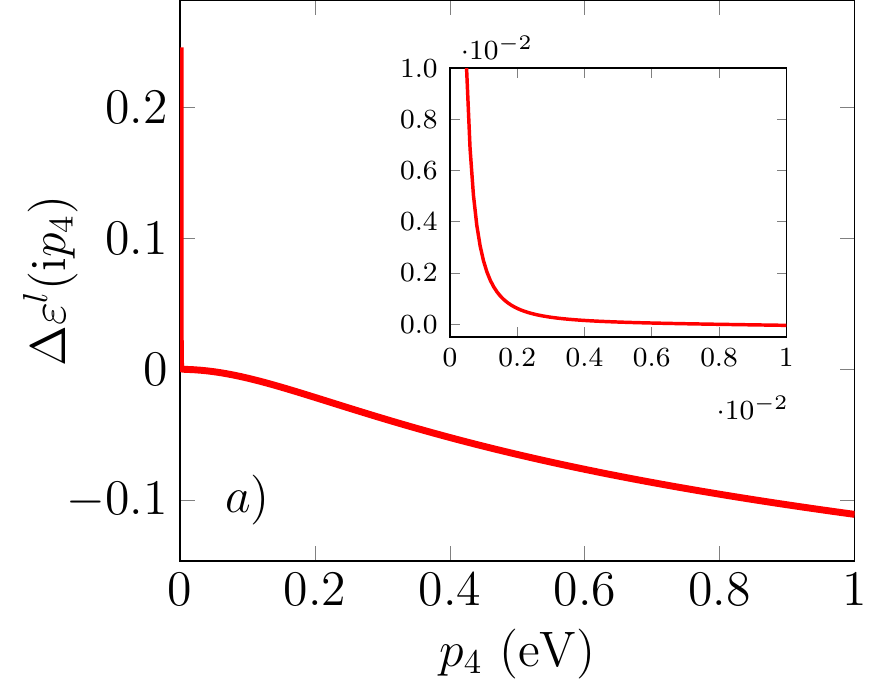}\includegraphics[width=5cm]{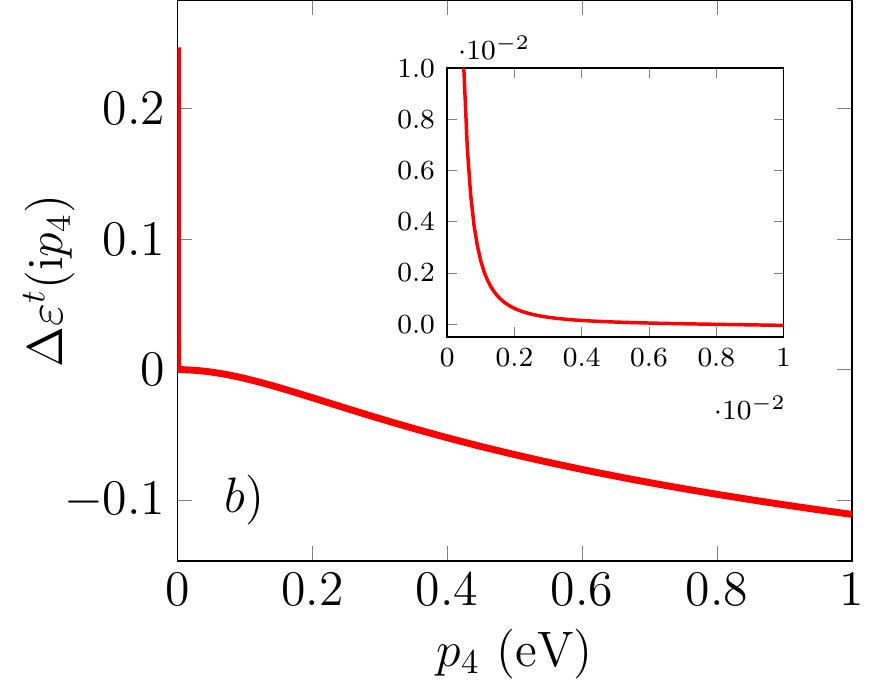}
	\caption{Total quantum correction $\Delta\varepsilon^{l,t} = \Delta_0\varepsilon^{l,t} + \Delta_T\varepsilon^{l,t}$ as functions of $p_4$ for $m=0.1$ eV, $|\vec{p}|=0.01$ eV, $T=100$ K, and $\mu = 0$. Figures (a) and (b) represent $\varepsilon^l$ and $\varepsilon^t$, respectively.}
	\label{fig:deltaepsilon}
\end{figure}

The other function $\varepsilon^t$ can be analysed along the same lines. However, due to the structure of reflection coefficients, see Eq. (\ref{rtetm}) below, the behaviour of $\varepsilon^t$ at small frequencies and momenta is less important. Typical plots for both $\varepsilon^l$ and $\varepsilon^t$ are given on Fig.\ \ref{fig:deltaepsilon}. Both graphs have very sharp positive peaks at low frequencies and long slowly-varying negative tails. The peak is so narrow that its influence is practically limited to the zeroth Matsubara frequency.\footnote{For $T=100$ K the first Matsubara frequency is $\xi_1=5.4\cdot 10^{-2}$ eV.}  

\section{Casimir interaction}\label{sec:Cas}
In this section, we study the Casimir interaction between two identical Dirac materials with parallel flat boundaries separated by a vacuum gap $a$. Let $x^3$ be the coordinate normal to the boundaries. The Casimir pressure (the force per unit area) is given by the celebrated Lifshitz formula
\begin{equation}
	P(a) = -\frac{T}{\pi} \sum_{n=0}^\infty{}' \int_0^\infty q_n p_\perp dp_\perp \sum_{l = \te,\tmm } \left(\frac{e^{2 a q_n}}{r^2_l(\ii \xi_n ,p_\perp)} - 1\right)^{-1}. \label{Lif}
\end{equation}
In this sum, the Euclidean frequency $p_4$ takes discrete Matsubara values
$p_4=\xi_n=2\pi n T$. The prime near summation symbol means that the term with $n=0$ enters with a factor of $1/2$. We also defined
\begin{equation}
p_\perp = \sqrt{p_1^2 + p_2^2},\qquad q_n = \sqrt{p_\perp^2 + \xi_n ^2}.
\end{equation}
In Eq.\ (\ref{Lif}), $r_\te$ and $r_\tmm$ are the reflection coefficients for TE and TM modes, respectively.

To compute the reflection coefficients one needs to know the behavior of the polarization tensor near the boundary. This can be done at least in principle by QFT methods as in \cite{Fialkovsky:2019rum,Kurkov:2020jet}. However, the computations done in these papers were quite complicated even for zero temperature and zero chemical potential. Thus, in the present work we ignore specific boundary contributions to the polarization tensor and adopt a much simpler approach 
\cite{Reuter:1948,Silin:1962,Kliewer:1968zz,Esquivel:2004zz} based on the assumption of specular reflection of charged particles at the boundary. In this approach, the reflection coefficients read
\begin{equation}
	r_\tmm  (\ii \xi_n ,p_\perp) = \frac{q_n - \xi_n Z_\tmm  (\ii \xi_n,p_\perp)}{q_n + \xi_n Z_\tmm  (\ii \xi_n,p_\perp)},\qquad \ r_\te (\ii \xi_n ,p_\perp) = \frac{q_n Z_\te (\ii \xi_n,p_\perp) - \xi_n}{q_n  Z_\te (\ii \xi_n,p_\perp) + \xi_n}.\label{rtetm}
\end{equation}
The surface impedance functions 
\begin{equation}\label{eq:Impedance}
	Z_\tmm  (\ii \xi_n,p_\perp) = \frac{\xi_n}{\pi} \int_{-\infty}^{+\infty} \frac{dp_3}{\vec{p}^{\,2}} \left(\frac{p_\perp^2}{\xi_n^2 \varepsilon_n^l} + \frac{p_3^2}{\vec{p}^{\,2} + \varepsilon_n^t \xi_n ^2}\right),\qquad Z_\te (\ii \xi_n,p_\perp) = \frac{\xi_n}{\pi} \int_{-\infty}^{+\infty} \frac{dp_3}{\vec{p}^{\,2} + \varepsilon_n^t \xi_n ^2},
\end{equation}
are completely defined by dielectric functions in the bulk. We used a shorthand notation $\varepsilon_n^{l,t}\equiv \varepsilon^{l,t} (\ii \xi_n,\vec{p})$.

As a reference point, we will use the Casimir pressure between two identical dielectrics with a constant permittivity $\varepsilon_0$, i.e. the materials where quantum corrections are neglected. The reflection coefficients in this case read
\begin{equation}
		r_\te^{(d)}(\ii \xi_n ,p_\perp) = \frac{q_n - \sqrt{p_\perp^2+ \varepsilon_0 \xi_n^2}}{q_n  + \sqrt{p_\perp^2+ \varepsilon_0 \xi_n^2}},\qquad 
		r_\tmm^{(d)}(\ii \xi_n ,p_\perp) = \frac{\varepsilon_0 q_n - \sqrt{p_\perp^2+ \varepsilon_0 \xi_n^2}}{\varepsilon_0 q_n + \sqrt{p_\perp^2+ \varepsilon_0 \xi_n^2}}. \label{rrd}
	\end{equation}

\begin{figure}[htb]
	\centering
	\includegraphics[width=4.9cm]{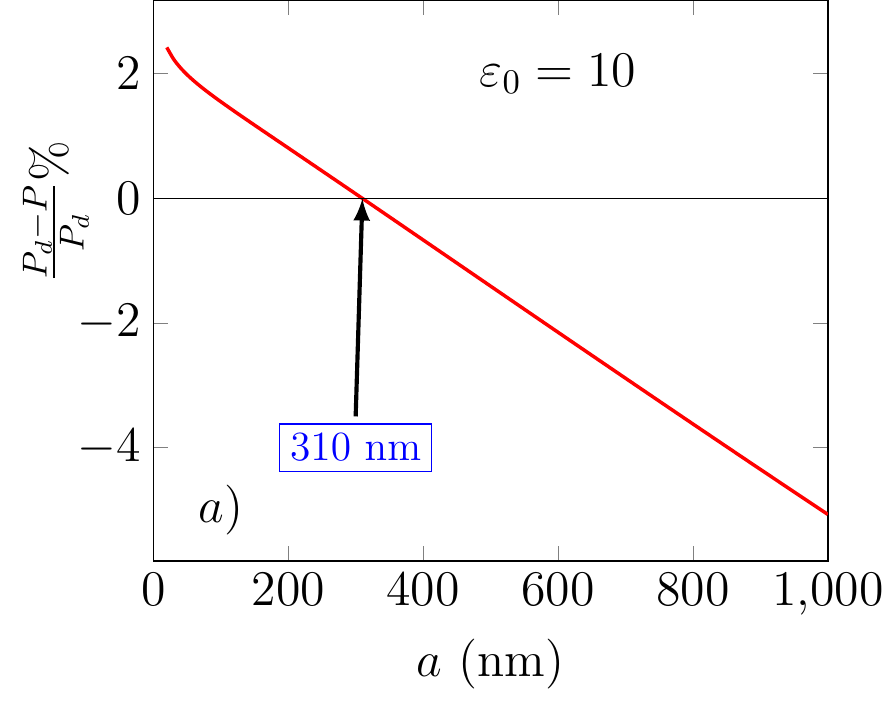}\includegraphics[width=5cm]{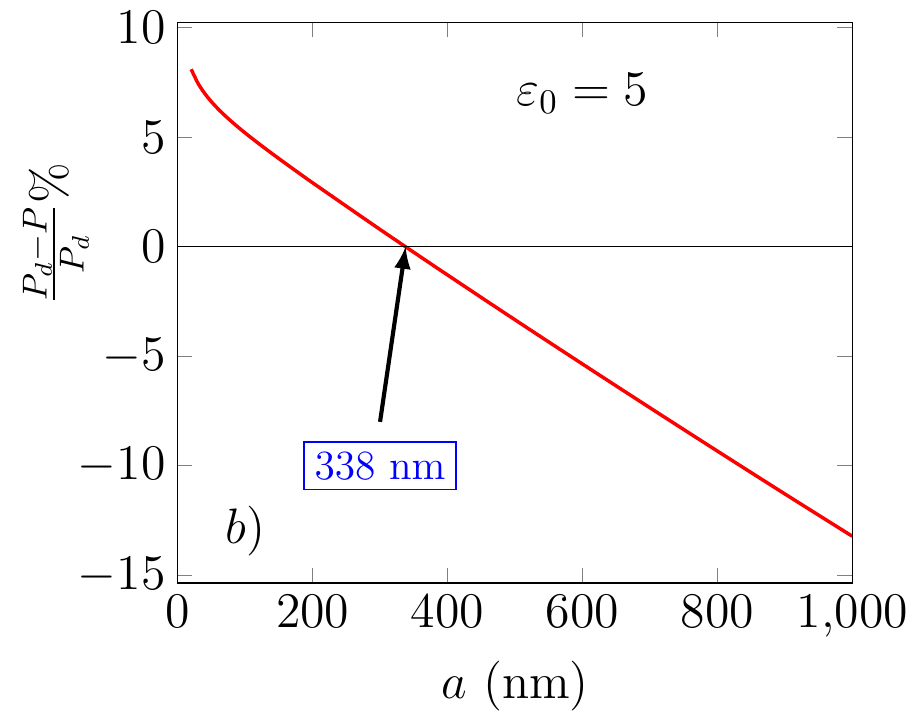}\includegraphics[width=5cm]{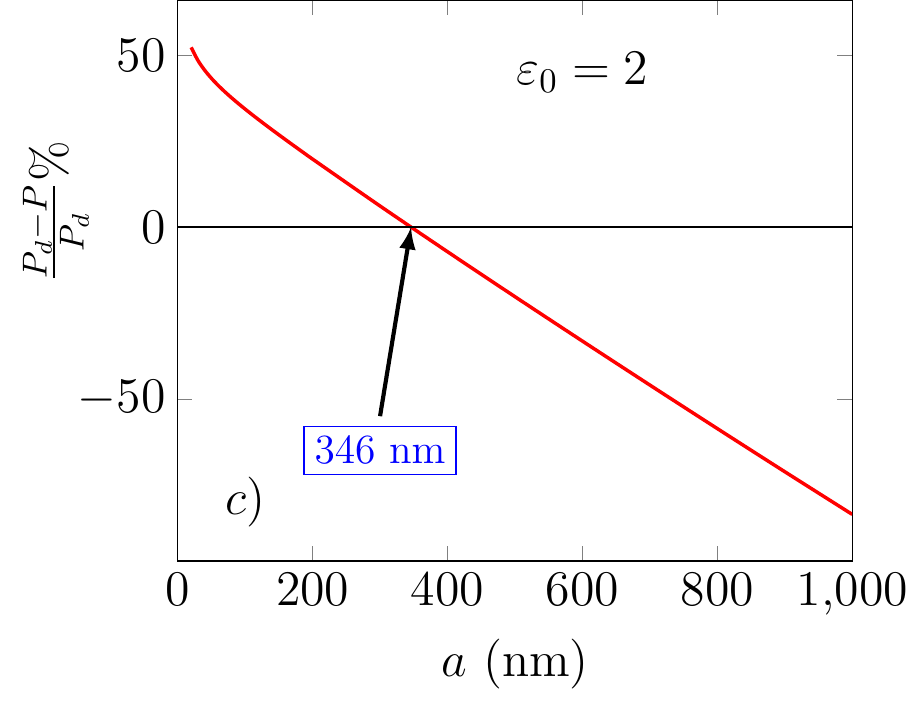}
	\caption{Relative variation of the Casimir pressure $(P_d - P)/P_d \% $ as a function of separation for $m=0.01$ eV, $T=100$ K, $\mu=0$ and three different values of $\varepsilon_0$: $\varepsilon_0=10$, $5$ and $2$ on Figs. (a), (b) and (c), respectively.}
	\label{fig:FF-8}
\end{figure}

\begin{figure}[htb]
	\centering
	\includegraphics[width=5cm]{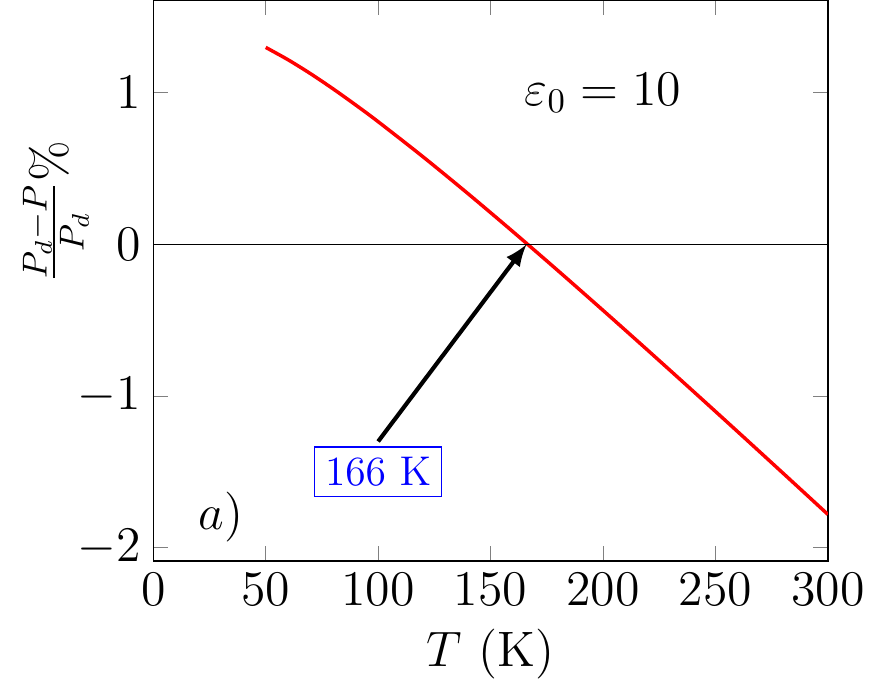}\includegraphics[width=5cm]{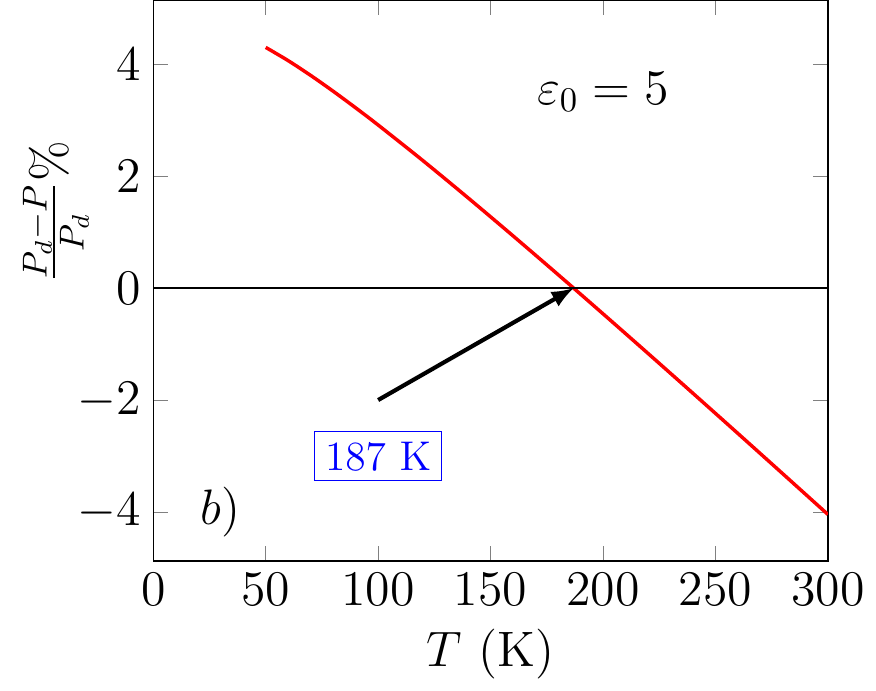}\includegraphics[width=5.1cm]{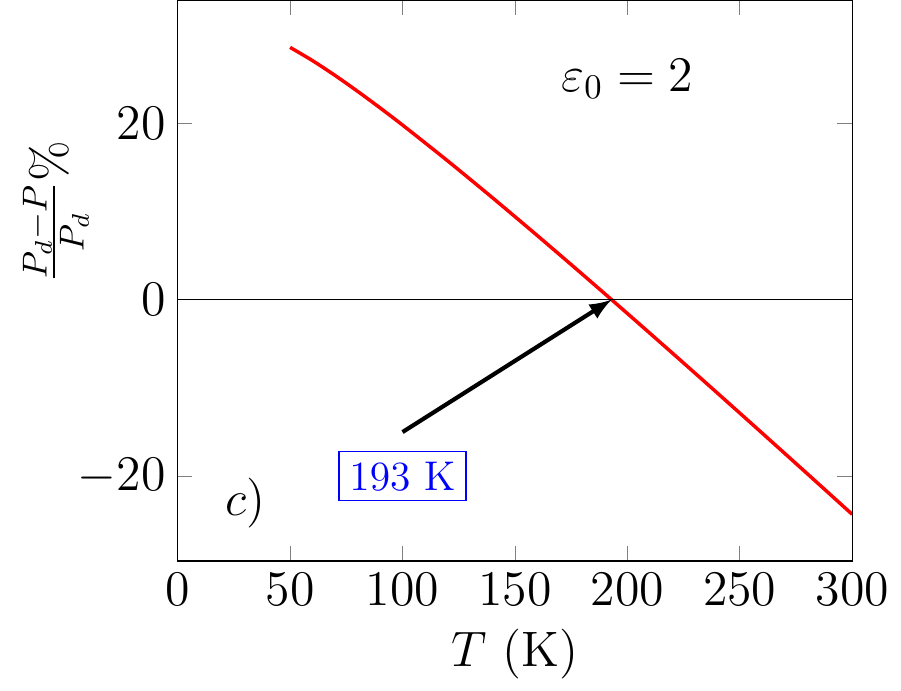}
	\caption{The relative difference in percent $(P_d - P)/P_d \% $. The values $m=0.01$ eV, $a=200$ nm, and $\mu=0$ eV are used. $\varepsilon_0$ is taken to be $10$, $5$, and $2$ at figures (a), (b) and (c), respectively.}
	\label{fig:FF-9}
\end{figure}

To estimate the effect of quantum corrections, let us consider the variation of Casimir pressure for a Dirac material $P$ relative to the Casimir pressure for dielectrics with a constant permittivity, denoted $P_d$, plotted as a function of the distance $a$, Fig. \ref{fig:FF-8}, and of the temperature $T$, Fig.\ \ref{fig:FF-9}. First of all, we see that this relative effect is tiny for $\varepsilon_0=10$, considerable for $\varepsilon_0=5$ and large for $\varepsilon_0=2$. However, this is just a background effect: the pressure $P_d$ used as a reference point increases as a function of $\varepsilon_0$. The shape of the functional dependence on $T$ is similar to that on $a$, which confirms the general observation on Casimir physics: the relevant parameter is the product $aT$. We see that at large distances and high temperatures quantum corrections increase the Casimir interaction, while for short distances and lower temperatures the interaction decreases. 

Qualitatively, the behavior of Casimir pressure depicted at Fig.\ \ref{fig:FF-8} and Fig.\ \ref{fig:FF-9} can be explained by the behavior of dielectric functions described in the previous section. Interestingly, the values of critical temperature, $T_c$, and critical distance, $a_c$, defined by $P_d=P$, have a rather weak dependence on $\varepsilon_0$. Thus, even though the Casimir interaction is a nonlinear effect, the presence of two characteristic regions may be explained by looking at the sign of quantum corrections without taking the value of $\varepsilon_0$ into account. It is very well know fact in the Casimir physics that the small $a$/low $T$ behavior of Casimir interaction is governed by the high-frequency behavior of reflection coefficients, while in the opposite limit the zeroth Matsubara frequency becomes increasingly important. This is roughly related to the presence of the damping factor $e^{2aq_n}$ in the Lifshitz formula (\ref{Lif}) and to the temperature dependence of the spacing in the Matsubara sum. Although the reflection coefficients are related to $\varepsilon^l$ and $\varepsilon^t$ through complicated formulae (\ref{rtetm}), one can deduce some qualitative results by looking at the properties of $\varepsilon^{l,t}$ which we discussed in the previous section. The zero-frequency positive peak in $\varepsilon^{l,t}$ effectively increases optical density of the material and thus leads to an increase of Casimir pressure at large $a$ and high $T$. The negative tail which becomes more visible at higher frequencies leads to a decrease of the Casimir pressure in the small $a$ and low $T$ regions. This is exactly what we see at Fig.\ \ref{fig:FF-8} and Fig.\ \ref{fig:FF-9}.  
 
\begin{figure}[h]
	\centering
	\includegraphics[width=5cm]{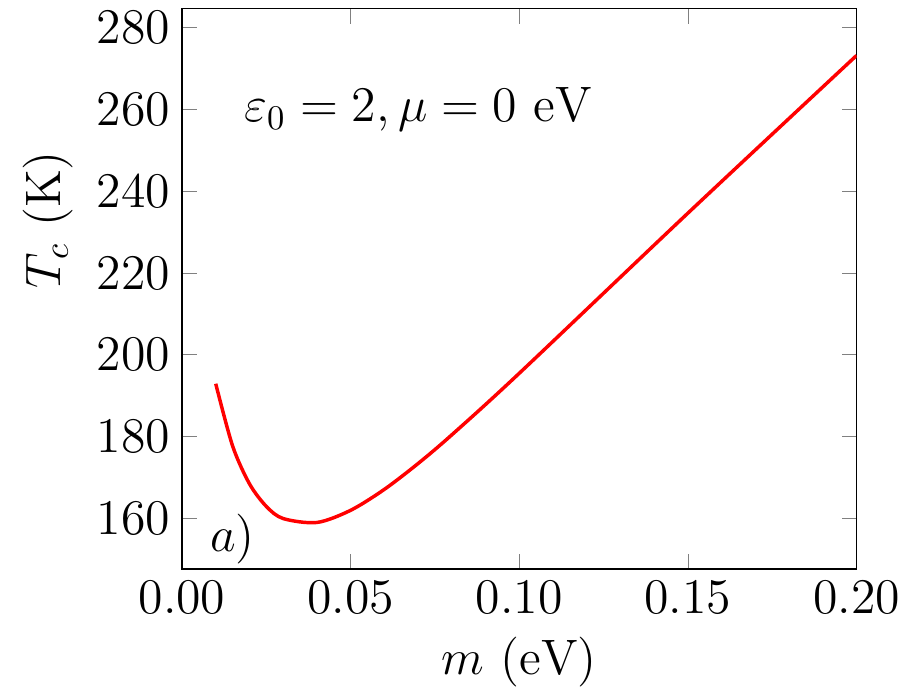}\includegraphics[width=5cm]{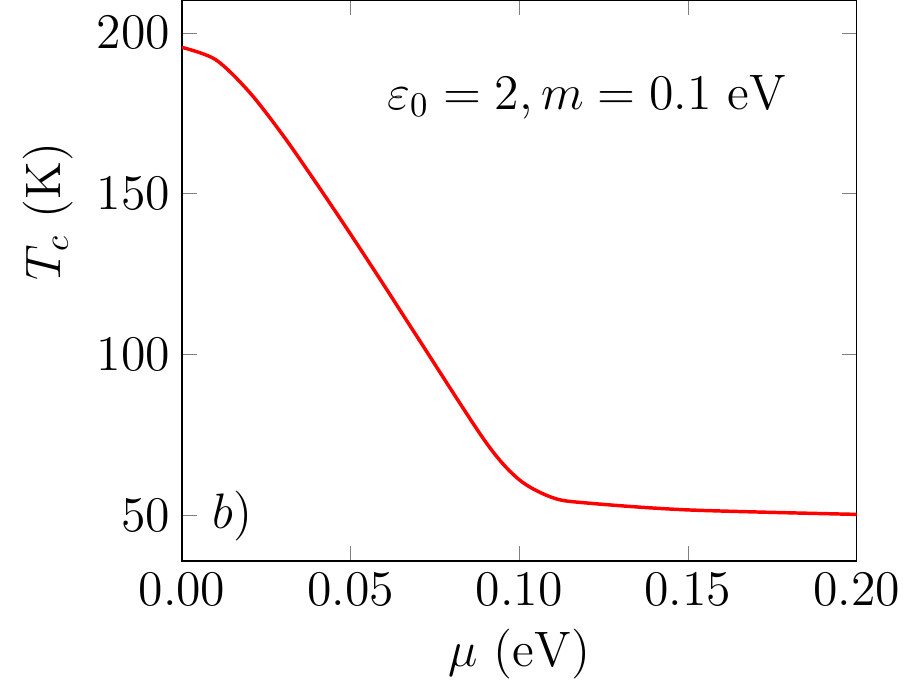}
	\caption{The dependence of critical temperature $T_{c}$ on the mass gap $m$ (Fig. (a)) and on the chemical potential $\mu$ (Fig. (b)). The values $\varepsilon_0=2$, $a=200$ nm are used.}
	\label{fig:FF-11}
\end{figure}
As another check of our qualitative understanding it is useful to consider the dependence of critical temperature $T_c$ (at which $P=P_d$) on the mass and chemical potential, see Fig.\ \ref{fig:FF-11}. It is clear on the basis of our arguments above, that the higher is the peak in $\varepsilon^{l,t}$ the smaller has to be $T_c$. From the expression (\ref{elmum}), we see that as long as $\mu<m$ the increase of $\mu$ leads to strong enhancement of $\varepsilon^l$ due an exponential factor. For $\mu>m$ the dependence of $\varepsilon^l$ on $\mu$ is weaker, as follows from (\ref{elmmu}). These two regions are clearly seen at Fig.\ \ref{fig:FF-11}(b). The dependence on $m$ is a bit more complicated. Actually, the increase of mass gap leads to a decrease of all quantum corrections, both in the "peak" and in the "tail". For $T\ll m$  suppression of the "peak" is exponential, see (\ref{elmum}) and is more important than suppression of the "tail". For lower values of $m$ the "tail" suppression wins. This latter effect cannot be confirmed by our analytic formulas, but it is clearly seen at Fig.\ \ref{fig:FF-11}(a). 

\begin{figure}[h]
	\centering
	\includegraphics[width=5cm]{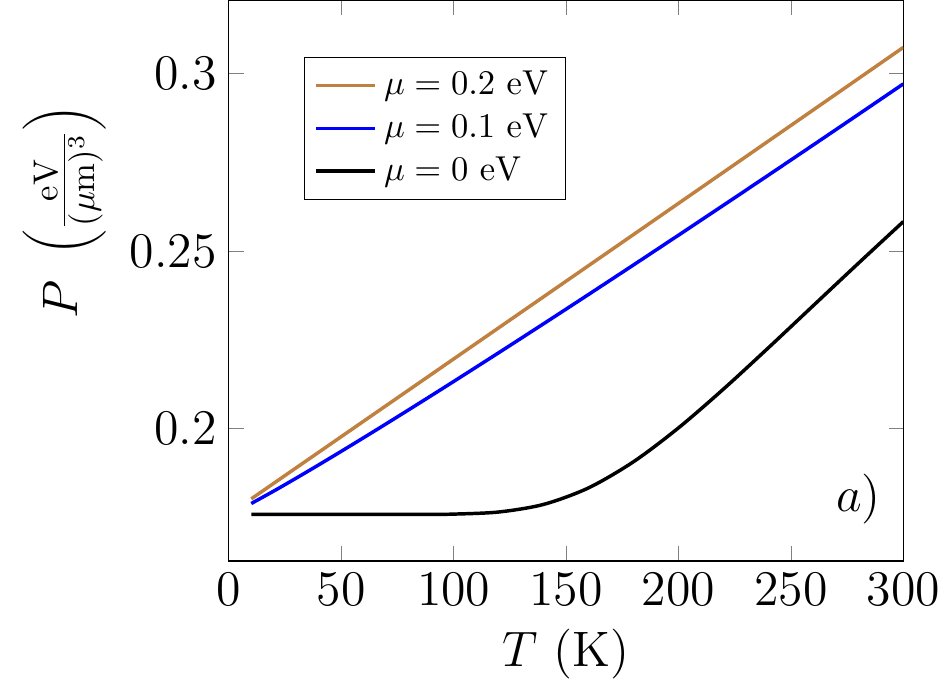}\includegraphics[width=5cm]{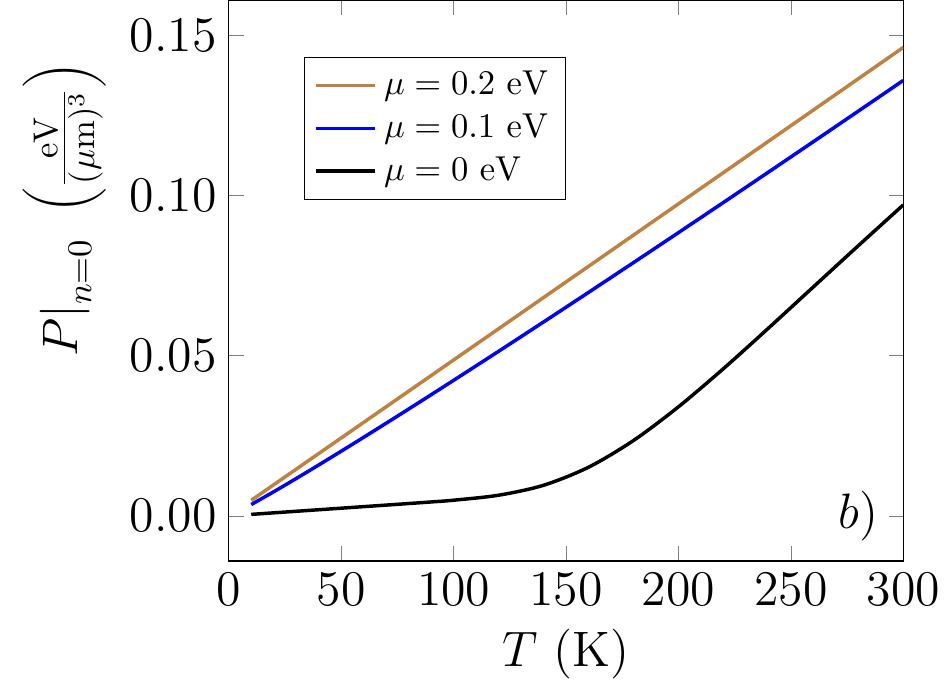}\includegraphics[width=4.8cm]{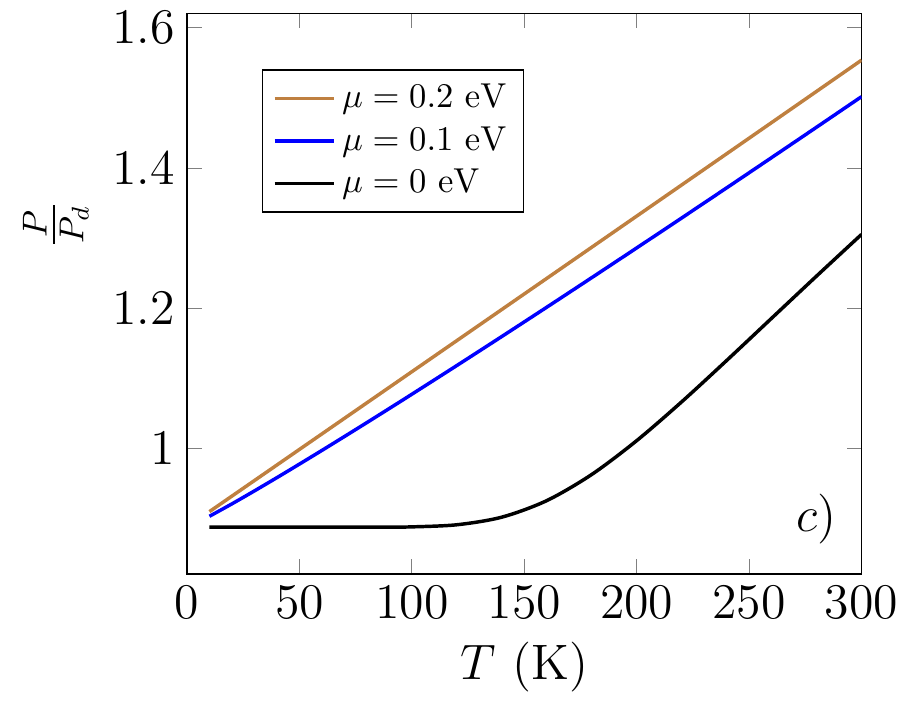}
	\caption{The plots of temperature dependence of (a) the Casimir pressure, (b) the contribution of zero Matsubara term to Casimir pressure, and (c) the relative Casimir pressure  for $\varepsilon_0 = 2$, $m=0.1$ eV, $a=200$ nm, and three different values of the chemical potential $\mu = 0,\ 0.1,\ 0.2$ eV.}
	\label{fig:difmu}
\end{figure}

Let us now discuss the contribution of zero Matsubara frequency term in (\ref{Lif}) to the Casimir pressure. By comparing the graphs on Fig.\ \ref{fig:difmu}(a) and on Fig.\ \ref{fig:difmu}(b) we observe that practically the whole dependence of Casimir pressure on $\mu$ and $T$ is due to this term. (A similar picture occurs until very small distances.) This is very remarkable fact which considerably simplifies numerical calculations. For $\mu=0$, and for small values of $\mu$ as well, there is no dependence of $P$ on $T$ for low temperature. This can be traced back to the strong temperature suppression in (\ref{elmum}). There is no such suppression for $\mu>m$, see (\ref{elmmu}). The $T$-dependence of pressure for $\mu>m$ is caused by an $O(T)$ correction which is not shown in (\ref{elmmu}). The behaviour of Casimir pressure at finite $T$ cannot be fully understood by asymptotic formulas. Thus, we have to rely on the numerics which shows that $P$ increases considerably with the increase of $T$ or $\mu$. A similar behaviour has been observed for the Casimir interaction of graphene \cite{Fialkovsky:2011pu,Bordag:2015zda}. We may conjecture that this property is  generic for all Dirac materials.


\section{Conclusions}\label{sec:con}
The purpose of this paper was to compute the bulk dielectric functions for Dirac materials at imaginary frequencies and to study their effect on the Casimir interaction. Let us summarize briefly our findings. The structure of dielectric functions was quite simple: a strong and sharp positive peak near the zero frequency and a long smooth negative tail, see Fig.\ \ref{fig:deltaepsilon}. This structure causes an increase of the Casimir pressure at large separations and high temperatures, and a decrease at small separation and low temperature, as compared to the interaction of dielectrics with a constant dielectric permittivity $\varepsilon_0$. Other properties of the Casimir pressure also got nice qualitative explanation through the properties of quantum corrections to dielectric functions. We found that the dependence of Casimir energy on temperature and chemical potential is almost entirely due to the contribution from the zeroth Matsubara term in the Lifshitz sum. This fact leads to great simplifications in the numerics. We like to stress that the effects of quantum corrections to dielectric functions are very well seen in the Casimir interaction, as well as the influence of mass gap and chemical potential.

In this work, boundary contributions to the polarization tensor have been totally neglected. To take them properly into account, one has to extend the results of \cite{Fialkovsky:2019rum,Kurkov:2020jet} to the case of non-zero temperature and chemical potential. We hope to address this problem in a future publication. The present work also did not consider possible contributions to the dielectric functions from other sources (from phonons, for example). Taking these contributions into account is also a task for some future work when we shall consider particular materials.

\begin{acknowledgments}
We are grateful to Maxim Kurkov for fruitful discussions. This work was supported in parts the project 2016/03319-6 of FAPESP, by the grants 305594/2019-2 and 428951/2018-0 of CNPq. Besides, D.V. was supported by the Tomsk State University Competitiveness Improvement Program. N.K. was supported in parts the project 2019/10719-9 of FAPESP and by the RFBR Projects 19-02-00496-a. 
\end{acknowledgments}

\appendix
\section{Computation of the polarization tensor}\label{sec:Com}
Here we sketch the computations of polarization tensor $\widehat{\Pi}^{\mu\nu}$ for  unit Fermi velocity, $v_F=1$. The formalism which we apply here is rather standard, see \cite{Shuryak:1980tp}. Exactly the same methods were uses in \cite{Bordag:2015rqa,Bordag:2015zda,Bordag:2015gla}, see also \cite{Falomir:2019yfk}, to compute the polarization tensor in graphene for non-zero temperature and chemical potential.

 After computing the traces of $\gamma$-matrices in (\ref{PiEu}) we arrive at the following expression (see notations below Eq.\ (\ref{vFPi}))
\begin{equation}
	\widehat{\Pi}^{\x} (p) = - 4 e^2 T \sum_{l=-\infty}^\infty \int \frac{d^3\vec{k}}{(2\pi)^3} \frac{Z^{\x}}{N}. \label{eq:Pi1}
\end{equation}
Here,
\begin{eqnarray}
	Z^{00} &=& m^2 - (k_4 -\ii \mu)(k_4 -\ii \mu + \ii p_0) + \vec{k}(\vec{k} - \vec{p}), \\
	Z^{\mathrm{tr}} &=& 4m^2 + 2 (k_4 -\ii \mu)(k_4 -\ii \mu + \ii p_0) + 2\vec{k}(\vec{k} - \vec{p}),\\
	N &=& [(k_4  - \ii \mu)^2 + E_k^2][(k_4 - \ii \mu + \ii p_0 )^2 + E_{k-p}^2 ] ,
\end{eqnarray}
with $k_0=\ii k_4=2\pi T(l+\tfrac 12)$, $E_k = \sqrt{\vec{k}^2 + m^2}$ and $E_{k-p} = \sqrt{(\vec{k}-\vec{p})^2 + m^2}$.

\begin{figure}
	\includegraphics[width=9cm]{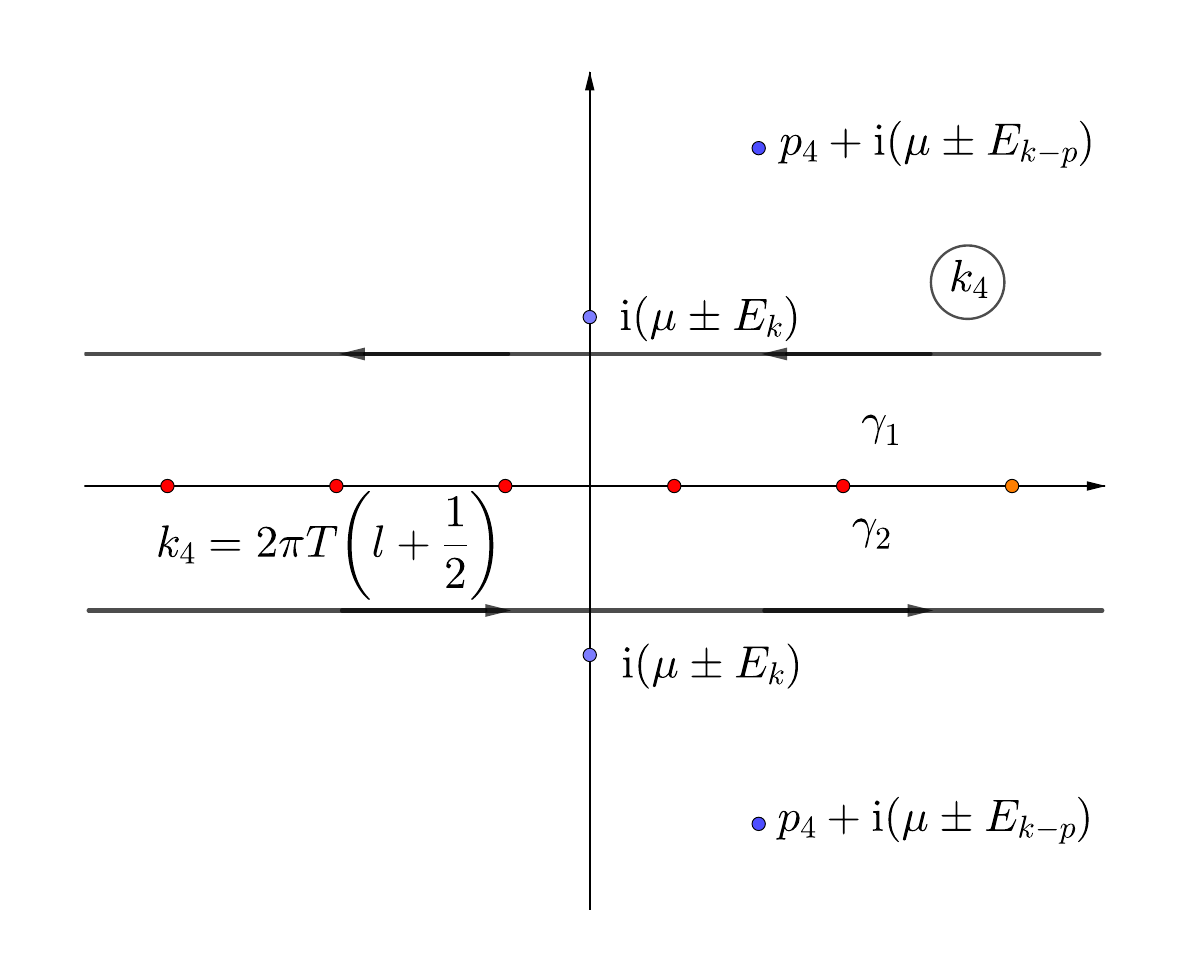}
	\caption{The integration contour $\gamma=\gamma_1\cup\gamma_2$ in the complex $k_4$ plane, see Eq.\ (\ref{eq:Pi2}). Here we also marked the positions of poles of the integrand. Depending on the signs of $\mu\pm E_k$ and of $\mu\pm E_{k-p}$ some poles may appear either in the upper or in the lower half-plane. Both possibilities are depicted. }\label{fig:cont}
\end{figure}

We express the sum over Matsubara frequencies through a contour integral 
\begin{equation}
	\widehat{\Pi}^{\x}(p) = 4e^2 \oint_{\gamma}  \frac{ dk_4}{1+e^{\ii \frac{k_4}{T}}}\int \frac{d^3\vec{k}}{(2\pi)^4} \frac{Z^{\x}}{N}. \label{eq:Pi2} 
\end{equation}
The contour $\gamma=\gamma_1\cup\gamma_2$ consists of two parts which are depicted at Fig.\ \ref{fig:cont}.

On the upper part $\gamma_1$ of the contour we use the identity
\begin{equation}
	\frac{1}{1+e^{\ii \frac{k_4}{T}}} = 1 - \frac{1}{1+e^{-\ii \frac{k_4}{T}}}.\label{decom}
\end{equation}
Contribution of the first (constant) term in (\ref{decom}) reads
\begin{equation}
\widehat\Pi_1^{\x}(p)=4e^2 \int_{+\infty}^{-\infty}dk_4\int\frac{d^3\vec{k}}{(2\pi)^4}\frac{Z^{\x}}{N} .\label{wPi1}
\end{equation}
With just the second term in (\ref{decom}) present under the integral, the integration contour $\gamma_1$ can be closed upwards. The contour $\gamma_2$ can be closed downwards. The integration over $k_4$ is done by computing the residues. After some long but otherwise straightforward algebra, we obtain the rest of
$\widehat{\Pi}_T=\widehat{\Pi}-\widehat{\Pi}_1$:
\begin{eqnarray}
	\widehat\Pi^{\x}_{T}(p)
	&=& \frac{e^2}{4\pi^3} \int \frac{d^3\vec{k}}{E_k } \left\{\frac{Z^{\x}[k_4=\ii (\mu + E_k)]}{E^2_{k-p} -(p_0 + E_k)^2} \left[n_\mu + n_{-\mu} - \theta (\mu - E_k) - \theta (-\mu - E_k)\right] \right. \nonumber \\
	&+&\left. \frac{Z^{\x}[k_4=\ii (\mu - E_k)]}{E^2_{k-p} -(p_0 - E_k)^2} \left[n_{-\mu} + n_{\mu} - \theta (\mu - E_k) - \theta ( -\mu - E_k) \right]\right\}.\label{PiT}
\end{eqnarray}
Here
\begin{equation}
	n_{\pm \mu} = \frac{1}{1+e^{\frac{E_k \pm \mu}{T}}}
\end{equation}
is the Boltzmann factor. We remind that $p_0=\ii p_4=2\pi\ii n T=\ii \xi_n$ is an imaginary bosonic Matsubara frequency. Thus, $n_{\mu\pm p_0}=n_\mu$.

It is useful to take the terms containing step functions $\theta$ out of the expressions (\ref{PiT}) and combine them with $\widehat{\Pi}_1$. The reshuffled contributions to polarization tensor read
\begin{eqnarray}
\widehat\Pi^{\x}_0 (p)  &=& -\frac{e^2}{4\pi^3}\int d^3\vec{k}  \left\{\int_{-\infty}^{+\infty}  \frac{dk_4}{\pi} \frac{Z^{\x}}{N} +  \frac{1}{E_k}\left[\frac{Z^{\x}[k_4=\ii (\mu + E_k)]}{E^2_{k-p} -(\ii p_4  + E_k)^2} + \frac{Z^{\x}[k_4=\ii (\mu - E_k)]}{E^2_{k-p} -(\ii p_4  - E_k)^2} \right] \theta (\mu^2 - E_k^2)\right\}, \label{eq:Pi0-1}\\
	\widehat\Pi^{\x}_{T,\mu} (p) &=& \frac{e^2}{4\pi^3} \int \frac{d^3\vec{k}}{E_k } \left\{\frac{Z^{\x}[k_4=\ii (\mu + E_k)]}{E^2_{k-p} -(\ii p_4  + E_k)^2}  + \frac{Z^{\x}[k_4=\ii (\mu - E_k)]}{E^2_{k-p} -(\ii p_4  - E_k)^2}\right\} \left(n_\mu + n_{-\mu} \right).\label{PiTm}
\end{eqnarray}
It remains to integrate over $k_4$ in (\ref{eq:Pi0-1}) and perform the angular integration in $d^3\vec{k}$ in both formulae above to obtain Eqs.\ (\ref{PiPP})--(\ref{eq:ISummary})  of the main text.

%


\begin{thebibliography}{34}%
\makeatletter
\providecommand \@ifxundefined [1]{%
 \@ifx{#1\undefined}
}%
\providecommand \@ifnum [1]{%
 \ifnum #1\expandafter \@firstoftwo
 \else \expandafter \@secondoftwo
 \fi
}%
\providecommand \@ifx [1]{%
 \ifx #1\expandafter \@firstoftwo
 \else \expandafter \@secondoftwo
 \fi
}%
\providecommand \natexlab [1]{#1}%
\providecommand \enquote  [1]{``#1''}%
\providecommand \bibnamefont  [1]{#1}%
\providecommand \bibfnamefont [1]{#1}%
\providecommand \citenamefont [1]{#1}%
\providecommand \href@noop [0]{\@secondoftwo}%
\providecommand \href [0]{\begingroup \@sanitize@url \@href}%
\providecommand \@href[1]{\@@startlink{#1}\@@href}%
\providecommand \@@href[1]{\endgroup#1\@@endlink}%
\providecommand \@sanitize@url [0]{\catcode `\\12\catcode `\$12\catcode
  `\&12\catcode `\#12\catcode `\^12\catcode `\_12\catcode `\%12\relax}%
\providecommand \@@startlink[1]{}%
\providecommand \@@endlink[0]{}%
\providecommand \url  [0]{\begingroup\@sanitize@url \@url }%
\providecommand \@url [1]{\endgroup\@href {#1}{\urlprefix }}%
\providecommand \urlprefix  [0]{URL }%
\providecommand \Eprint [0]{\href }%
\providecommand \doibase [0]{https://doi.org/}%
\providecommand \selectlanguage [0]{\@gobble}%
\providecommand \bibinfo  [0]{\@secondoftwo}%
\providecommand \bibfield  [0]{\@secondoftwo}%
\providecommand \translation [1]{[#1]}%
\providecommand \BibitemOpen [0]{}%
\providecommand \bibitemStop [0]{}%
\providecommand \bibitemNoStop [0]{.\EOS\space}%
\providecommand \EOS [0]{\spacefactor3000\relax}%
\providecommand \BibitemShut  [1]{\csname bibitem#1\endcsname}%
\let\auto@bib@innerbib\@empty
\bibitem [{\citenamefont {Qi}\ and\ \citenamefont {Zhang}(2011)}]{Qi:2011zya}%
  \BibitemOpen
  \bibfield  {author} {\bibinfo {author} {\bibfnamefont {X.~L.}\ \bibnamefont
  {Qi}}\ and\ \bibinfo {author} {\bibfnamefont {S.~C.}\ \bibnamefont {Zhang}},\
  }\bibfield  {title} {\bibinfo {title} {{Topological insulators and
  superconductors}},\ }\href {https://doi.org/10.1103/RevModPhys.83.1057}
  {\bibfield  {journal} {\bibinfo  {journal} {Rev. Mod. Phys.}\ }\textbf
  {\bibinfo {volume} {83}},\ \bibinfo {pages} {1057} (\bibinfo {year}
  {2011})},\ \Eprint {https://arxiv.org/abs/1008.2026} {arXiv:1008.2026
  [cond-mat.mes-hall]} \BibitemShut {NoStop}%
\bibitem [{\citenamefont {Hasan}\ and\ \citenamefont
  {Kane}(2010)}]{Hasan:2010xy}%
  \BibitemOpen
  \bibfield  {author} {\bibinfo {author} {\bibfnamefont {M.~Z.}\ \bibnamefont
  {Hasan}}\ and\ \bibinfo {author} {\bibfnamefont {C.~L.}\ \bibnamefont
  {Kane}},\ }\bibfield  {title} {\bibinfo {title} {{Topological Insulators}},\
  }\href {https://doi.org/10.1103/RevModPhys.82.3045} {\bibfield  {journal}
  {\bibinfo  {journal} {Rev. Mod. Phys.}\ }\textbf {\bibinfo {volume} {82}},\
  \bibinfo {pages} {3045} (\bibinfo {year} {2010})},\ \Eprint
  {https://arxiv.org/abs/1002.3895} {arXiv:1002.3895 [cond-mat.mes-hall]}
  \BibitemShut {NoStop}%
\bibitem [{\citenamefont {Bordag}\ \emph
  {et~al.}(2009{\natexlab{a}})\citenamefont {Bordag}, \citenamefont
  {Klimchitskaya}, \citenamefont {Mohideen},\ and\ \citenamefont
  {Mostepanenko}}]{Bordag:2009zzd}%
  \BibitemOpen
  \bibfield  {author} {\bibinfo {author} {\bibfnamefont {M.}~\bibnamefont
  {Bordag}}, \bibinfo {author} {\bibfnamefont {G.~L.}\ \bibnamefont
  {Klimchitskaya}}, \bibinfo {author} {\bibfnamefont {U.}~\bibnamefont
  {Mohideen}},\ and\ \bibinfo {author} {\bibfnamefont {V.~M.}\ \bibnamefont
  {Mostepanenko}},\ }\href
  {https://doi.org/10.1093/acprof:oso/9780199238743.001.0001} {\emph {\bibinfo
  {title} {{Advances in the Casimir effect}}}},\ \bibinfo {series} {Int. Ser.
  Monogr. Phys.}, Vol.\ \bibinfo {volume} {145}\ (\bibinfo  {publisher}
  {{Oxford University Press, Oxford, UK}},\ \bibinfo {year} {2009})\ pp.\
  \bibinfo {pages} {1--768}\BibitemShut {NoStop}%
\bibitem [{\citenamefont {Woods}\ \emph {et~al.}(2016)\citenamefont {Woods},
  \citenamefont {Dalvit}, \citenamefont {Tkatchenko}, \citenamefont
  {Rodriguez-Lopez}, \citenamefont {Rodriguez},\ and\ \citenamefont
  {Podgornik}}]{Woods:2015pla}%
  \BibitemOpen
  \bibfield  {author} {\bibinfo {author} {\bibfnamefont {L.}~\bibnamefont
  {Woods}}, \bibinfo {author} {\bibfnamefont {D.}~\bibnamefont {Dalvit}},
  \bibinfo {author} {\bibfnamefont {A.}~\bibnamefont {Tkatchenko}}, \bibinfo
  {author} {\bibfnamefont {P.}~\bibnamefont {Rodriguez-Lopez}}, \bibinfo
  {author} {\bibfnamefont {A.}~\bibnamefont {Rodriguez}},\ and\ \bibinfo
  {author} {\bibfnamefont {R.}~\bibnamefont {Podgornik}},\ }\bibfield  {title}
  {\bibinfo {title} {{Materials perspective on Casimir and van der Waals
  interactions}},\ }\href {https://doi.org/10.1103/RevModPhys.88.45003,
  10.1103/RevModPhys.88.045003} {\bibfield  {journal} {\bibinfo  {journal}
  {Rev. Mod. Phys.}\ }\textbf {\bibinfo {volume} {88}},\ \bibinfo {pages}
  {045003} (\bibinfo {year} {2016})},\ \Eprint
  {https://arxiv.org/abs/1509.03338} {arXiv:1509.03338 [cond-mat.mtrl-sci]}
  \BibitemShut {NoStop}%
\bibitem [{\citenamefont {Lu}(2021)}]{Lu:2021jvu}%
  \BibitemOpen
  \bibfield  {author} {\bibinfo {author} {\bibfnamefont {B.-S.}\ \bibnamefont
  {Lu}},\ }\bibfield  {title} {\bibinfo {title} {The casimir effect in
  topological matter},\ }\href {https://doi.org/10.3390/universe7070237}
  {\bibfield  {journal} {\bibinfo  {journal} {Universe}\ }\textbf {\bibinfo
  {volume} {7}},\ \bibinfo {pages} {237} (\bibinfo {year} {2021})}\BibitemShut
  {NoStop}%
\bibitem [{\citenamefont {Rodriguez-Lopez}\ \emph {et~al.}(2020)\citenamefont
  {Rodriguez-Lopez}, \citenamefont {Popescu}, \citenamefont {Fialkovsky},
  \citenamefont {Khusnutdinov},\ and\ \citenamefont
  {Woods}}]{Rodriguez-Lopez:2020:scorCitIaIIWs}%
  \BibitemOpen
  \bibfield  {author} {\bibinfo {author} {\bibfnamefont {P.}~\bibnamefont
  {Rodriguez-Lopez}}, \bibinfo {author} {\bibfnamefont {A.}~\bibnamefont
  {Popescu}}, \bibinfo {author} {\bibfnamefont {I.}~\bibnamefont {Fialkovsky}},
  \bibinfo {author} {\bibfnamefont {N.}~\bibnamefont {Khusnutdinov}},\ and\
  \bibinfo {author} {\bibfnamefont {L.~M.}\ \bibnamefont {Woods}},\ }\bibfield
  {title} {\bibinfo {title} {{Signatures of complex optical response in Casimir
  interactions of type I and II Weyl semimetals}},\ }\href
  {https://doi.org/10.1038/s43246-020-0015-4} {\bibfield  {journal} {\bibinfo
  {journal} {Commun. Mater.}\ }\textbf {\bibinfo {volume} {1}},\ \bibinfo
  {pages} {14} (\bibinfo {year} {2020})}\BibitemShut {NoStop}%
\bibitem [{\citenamefont {Farias}\ \emph {et~al.}(2020)\citenamefont {Farias},
  \citenamefont {Zyuzin},\ and\ \citenamefont {Schmidt}}]{Farias:2020qqp}%
  \BibitemOpen
  \bibfield  {author} {\bibinfo {author} {\bibfnamefont {M.~B.}\ \bibnamefont
  {Farias}}, \bibinfo {author} {\bibfnamefont {A.~A.}\ \bibnamefont {Zyuzin}},\
  and\ \bibinfo {author} {\bibfnamefont {T.~L.}\ \bibnamefont {Schmidt}},\
  }\bibfield  {title} {\bibinfo {title} {{Casimir force between Weyl semimetals
  in a chiral medium}},\ }\href {https://doi.org/10.1103/PhysRevB.101.235446}
  {\bibfield  {journal} {\bibinfo  {journal} {Phys. Rev. B}\ }\textbf {\bibinfo
  {volume} {101}},\ \bibinfo {pages} {235446} (\bibinfo {year} {2020})},\
  \Eprint {https://arxiv.org/abs/2001.10329} {arXiv:2001.10329
  [cond-mat.mes-hall]} \BibitemShut {NoStop}%
\bibitem [{\citenamefont {Ishikawa}\ \emph {et~al.}(2021)\citenamefont
  {Ishikawa}, \citenamefont {Nakayama},\ and\ \citenamefont
  {Suzuki}}]{Ishikawa:2020icy}%
  \BibitemOpen
  \bibfield  {author} {\bibinfo {author} {\bibfnamefont {T.}~\bibnamefont
  {Ishikawa}}, \bibinfo {author} {\bibfnamefont {K.}~\bibnamefont {Nakayama}},\
  and\ \bibinfo {author} {\bibfnamefont {K.}~\bibnamefont {Suzuki}},\
  }\bibfield  {title} {\bibinfo {title} {{Lattice-fermionic Casimir effect and
  topological insulators}},\ }\href
  {https://doi.org/10.1103/PhysRevResearch.3.023201} {\bibfield  {journal}
  {\bibinfo  {journal} {Phys. Rev. Res.}\ }\textbf {\bibinfo {volume} {3}},\
  \bibinfo {pages} {023201} (\bibinfo {year} {2021})},\ \Eprint
  {https://arxiv.org/abs/2012.11398} {arXiv:2012.11398 [hep-lat]} \BibitemShut
  {NoStop}%
\bibitem [{\citenamefont {Bordag}\ \emph
  {et~al.}(2009{\natexlab{b}})\citenamefont {Bordag}, \citenamefont
  {Fialkovsky}, \citenamefont {Gitman},\ and\ \citenamefont
  {Vassilevich}}]{Bordag:2009fz}%
  \BibitemOpen
  \bibfield  {author} {\bibinfo {author} {\bibfnamefont {M.}~\bibnamefont
  {Bordag}}, \bibinfo {author} {\bibfnamefont {I.~V.}\ \bibnamefont
  {Fialkovsky}}, \bibinfo {author} {\bibfnamefont {D.~M.}\ \bibnamefont
  {Gitman}},\ and\ \bibinfo {author} {\bibfnamefont {D.~V.}\ \bibnamefont
  {Vassilevich}},\ }\bibfield  {title} {\bibinfo {title} {{Casimir interaction
  between a perfect conductor and graphene described by the Dirac model}},\
  }\href {https://doi.org/10.1103/PhysRevB.80.245406} {\bibfield  {journal}
  {\bibinfo  {journal} {Phys. Rev.}\ }\textbf {\bibinfo {volume} {B 80}},\
  \bibinfo {pages} {245406} (\bibinfo {year} {2009}{\natexlab{b}})},\ \Eprint
  {https://arxiv.org/abs/0907.3242} {arXiv:0907.3242 [hep-th]} \BibitemShut
  {NoStop}%
\bibitem [{\citenamefont {Fialkovsky}\ \emph {et~al.}(2011)\citenamefont
  {Fialkovsky}, \citenamefont {Marachevsky},\ and\ \citenamefont
  {Vassilevich}}]{Fialkovsky:2011pu}%
  \BibitemOpen
  \bibfield  {author} {\bibinfo {author} {\bibfnamefont {I.~V.}\ \bibnamefont
  {Fialkovsky}}, \bibinfo {author} {\bibfnamefont {V.~N.}\ \bibnamefont
  {Marachevsky}},\ and\ \bibinfo {author} {\bibfnamefont {D.~V.}\ \bibnamefont
  {Vassilevich}},\ }\bibfield  {title} {\bibinfo {title} {{Finite temperature
  Casimir effect for graphene}},\ }\href
  {https://doi.org/10.1103/PhysRevB.84.035446} {\bibfield  {journal} {\bibinfo
  {journal} {Phys. Rev.}\ }\textbf {\bibinfo {volume} {B 84}},\ \bibinfo
  {pages} {035446} (\bibinfo {year} {2011})},\ \Eprint
  {https://arxiv.org/abs/1102.1757} {arXiv:1102.1757 [hep-th]} \BibitemShut
  {NoStop}%
\bibitem [{\citenamefont {Banishev}\ \emph {et~al.}(2013)\citenamefont
  {Banishev}, \citenamefont {Wen}, \citenamefont {Kawakami}, \citenamefont
  {Klimchitskaya}, \citenamefont {Mostepanenko},\ and\ \citenamefont
  {Mohideen}}]{Banishev:2013}%
  \BibitemOpen
  \bibfield  {author} {\bibinfo {author} {\bibfnamefont {A.~A.}\ \bibnamefont
  {Banishev}}, \bibinfo {author} {\bibfnamefont {H.}~\bibnamefont {Wen}},
  \bibinfo {author} {\bibfnamefont {R.~K.}\ \bibnamefont {Kawakami}}, \bibinfo
  {author} {\bibfnamefont {G.~L.}\ \bibnamefont {Klimchitskaya}}, \bibinfo
  {author} {\bibfnamefont {V.~M.}\ \bibnamefont {Mostepanenko}},\ and\ \bibinfo
  {author} {\bibfnamefont {U.}~\bibnamefont {Mohideen}},\ }\bibfield  {title}
  {\bibinfo {title} {{Measuring the Casimir force gradient from graphene on a
  SiO$_2$ substrate}},\ }\href {https://doi.org/10.1103/PhysRevB.87.205433}
  {\bibfield  {journal} {\bibinfo  {journal} {Phys. Rev.}\ }\textbf {\bibinfo
  {volume} {B 87}},\ \bibinfo {pages} {205433} (\bibinfo {year}
  {2013})}\BibitemShut {NoStop}%
\bibitem [{\citenamefont {Klimchitskaya}\ \emph {et~al.}(2014)\citenamefont
  {Klimchitskaya}, \citenamefont {Mohideen},\ and\ \citenamefont
  {Mostepanenko}}]{Klimchitskaya:2014axa}%
  \BibitemOpen
  \bibfield  {author} {\bibinfo {author} {\bibfnamefont {G.~L.}\ \bibnamefont
  {Klimchitskaya}}, \bibinfo {author} {\bibfnamefont {U.}~\bibnamefont
  {Mohideen}},\ and\ \bibinfo {author} {\bibfnamefont {V.~M.}\ \bibnamefont
  {Mostepanenko}},\ }\bibfield  {title} {\bibinfo {title} {{Theory of the
  Casimir interaction from graphene-coated substrates using the polarization
  tensor and comparison with experiment}},\ }\href
  {https://doi.org/10.1103/PhysRevB.89.115419} {\bibfield  {journal} {\bibinfo
  {journal} {Phys. Rev.}\ }\textbf {\bibinfo {volume} {B 89}},\ \bibinfo
  {pages} {115419} (\bibinfo {year} {2014})}\BibitemShut {NoStop}%
\bibitem [{\citenamefont {Liu}\ \emph {et~al.}(2021)\citenamefont {Liu},
  \citenamefont {Zhang}, \citenamefont {Klimchitskaya}, \citenamefont
  {Mostepanenko},\ and\ \citenamefont {Mohideen}}]{Liu:2021ice}%
  \BibitemOpen
  \bibfield  {author} {\bibinfo {author} {\bibfnamefont {M.}~\bibnamefont
  {Liu}}, \bibinfo {author} {\bibfnamefont {Y.}~\bibnamefont {Zhang}}, \bibinfo
  {author} {\bibfnamefont {G.~L.}\ \bibnamefont {Klimchitskaya}}, \bibinfo
  {author} {\bibfnamefont {V.~M.}\ \bibnamefont {Mostepanenko}},\ and\ \bibinfo
  {author} {\bibfnamefont {U.}~\bibnamefont {Mohideen}},\ }\bibfield  {title}
  {\bibinfo {title} {{Demonstration of an Unusual Thermal Effect in the Casimir
  Force from Graphene}},\ }\href
  {https://doi.org/10.1103/PhysRevLett.126.206802} {\bibfield  {journal}
  {\bibinfo  {journal} {Phys. Rev. Lett.}\ }\textbf {\bibinfo {volume} {126}},\
  \bibinfo {pages} {206802} (\bibinfo {year} {2021})},\ \Eprint
  {https://arxiv.org/abs/2104.13598} {arXiv:2104.13598 [quant-ph]} \BibitemShut
  {NoStop}%
\bibitem [{\citenamefont {Klimchitskaya}\ and\ \citenamefont
  {Mostepanenko}(2020)}]{Klimchitskaya:2020qmy}%
  \BibitemOpen
  \bibfield  {author} {\bibinfo {author} {\bibfnamefont {G.~L.}\ \bibnamefont
  {Klimchitskaya}}\ and\ \bibinfo {author} {\bibfnamefont {V.~M.}\ \bibnamefont
  {Mostepanenko}},\ }\bibfield  {title} {\bibinfo {title} {{An alternative
  response to the off-shell quantum fluctuations: a step forward in resolution
  of the Casimir puzzle}},\ }\href
  {https://doi.org/10.1140/epjc/s10052-020-08465-y} {\bibfield  {journal}
  {\bibinfo  {journal} {Eur. Phys. J.}\ }\textbf {\bibinfo {volume} {C80}},\
  \bibinfo {pages} {900} (\bibinfo {year} {2020})},\ \Eprint
  {https://arxiv.org/abs/2010.00998} {arXiv:2010.00998 [quant-ph]} \BibitemShut
  {NoStop}%
\bibitem [{\citenamefont {Klimchitskaya}\ and\ \citenamefont
  {Mostepanenko}(2021)}]{Klimchitskaya:2021qri}%
  \BibitemOpen
  \bibfield  {author} {\bibinfo {author} {\bibfnamefont {G.~L.}\ \bibnamefont
  {Klimchitskaya}}\ and\ \bibinfo {author} {\bibfnamefont {V.~M.}\ \bibnamefont
  {Mostepanenko}},\ }\bibfield  {title} {\bibinfo {title} {{Casimir entropy and
  nonlocal response functions to the off-shell quantum fluctuations}},\ }\href
  {https://doi.org/10.1103/PhysRevD.103.096007} {\bibfield  {journal} {\bibinfo
   {journal} {Phys. Rev.}\ }\textbf {\bibinfo {volume} {D103}},\ \bibinfo
  {pages} {096007} (\bibinfo {year} {2021})},\ \Eprint
  {https://arxiv.org/abs/2104.06351} {arXiv:2104.06351 [quant-ph]} \BibitemShut
  {NoStop}%
\bibitem [{\citenamefont {Lindhard}(1954)}]{Lindhard:1954}%
  \BibitemOpen
  \bibfield  {author} {\bibinfo {author} {\bibfnamefont {J.}~\bibnamefont
  {Lindhard}},\ }\bibfield  {title} {\bibinfo {title} {On the properties of a
  gas of charged particles},\ }\href@noop {} {\bibfield  {journal} {\bibinfo
  {journal} {Dan. Mat. Fys. Medd.}\ }\textbf {\bibinfo {volume} {28}},\
  \bibinfo {pages} {2} (\bibinfo {year} {1954})}\BibitemShut {NoStop}%
\bibitem [{\citenamefont {Bordag}\ and\ \citenamefont
  {Pirozhenko}(2018)}]{bord18-10-74}%
  \BibitemOpen
  \bibfield  {author} {\bibinfo {author} {\bibfnamefont {M.}~\bibnamefont
  {Bordag}}\ and\ \bibinfo {author} {\bibfnamefont {I.}~\bibnamefont
  {Pirozhenko}},\ }\bibfield  {title} {\bibinfo {title} {Dispersion forces
  between fields confined to half spaces},\ }\href
  {https://doi.org/10.3390/sym10030074} {\bibfield  {journal} {\bibinfo
  {journal} {Symmetry}\ }\textbf {\bibinfo {volume} {10}},\ \bibinfo {pages}
  {74} (\bibinfo {year} {2018})},\ \Eprint
  {https://arxiv.org/abs/quant-ph/1803.08113} {arXiv:quant-ph/1803.08113
  [quant-ph]} \BibitemShut {NoStop}%
\bibitem [{\citenamefont {Fialkovsky}\ \emph {et~al.}(2019)\citenamefont
  {Fialkovsky}, \citenamefont {Kurkov},\ and\ \citenamefont
  {Vassilevich}}]{Fialkovsky:2019rum}%
  \BibitemOpen
  \bibfield  {author} {\bibinfo {author} {\bibfnamefont {I.}~\bibnamefont
  {Fialkovsky}}, \bibinfo {author} {\bibfnamefont {M.}~\bibnamefont {Kurkov}},\
  and\ \bibinfo {author} {\bibfnamefont {D.}~\bibnamefont {Vassilevich}},\
  }\bibfield  {title} {\bibinfo {title} {{Quantum Dirac fermions in a
  half-space and their interaction with an electromagnetic field}},\ }\href
  {https://doi.org/10.1103/PhysRevD.100.045026} {\bibfield  {journal} {\bibinfo
   {journal} {Phys. Rev.}\ }\textbf {\bibinfo {volume} {D100}},\ \bibinfo
  {pages} {045026} (\bibinfo {year} {2019})},\ \Eprint
  {https://arxiv.org/abs/1906.06704} {arXiv:1906.06704 [hep-th]} \BibitemShut
  {NoStop}%
\bibitem [{\citenamefont {Isobe}\ and\ \citenamefont
  {Nagaosa}(2012)}]{Isobe:2012vh}%
  \BibitemOpen
  \bibfield  {author} {\bibinfo {author} {\bibfnamefont {H.}~\bibnamefont
  {Isobe}}\ and\ \bibinfo {author} {\bibfnamefont {N.}~\bibnamefont
  {Nagaosa}},\ }\bibfield  {title} {\bibinfo {title} {{Theory of quantum
  critical phenomenon in topological insulator - (3+1)D quantum electrodynamics
  in solids -}},\ }\href {https://doi.org/10.1103/PhysRevB.86.165127}
  {\bibfield  {journal} {\bibinfo  {journal} {Phys. Rev.}\ }\textbf {\bibinfo
  {volume} {B86}},\ \bibinfo {pages} {165127} (\bibinfo {year} {2012})},\
  \Eprint {https://arxiv.org/abs/1205.2427} {arXiv:1205.2427
  [cond-mat.mes-hall]} \BibitemShut {NoStop}%
\bibitem [{\citenamefont {Roy}\ \emph {et~al.}(2016)\citenamefont {Roy},
  \citenamefont {Juricic},\ and\ \citenamefont {Herbut}}]{Roy:2015zna}%
  \BibitemOpen
  \bibfield  {author} {\bibinfo {author} {\bibfnamefont {B.}~\bibnamefont
  {Roy}}, \bibinfo {author} {\bibfnamefont {V.}~\bibnamefont {Juricic}},\ and\
  \bibinfo {author} {\bibfnamefont {I.~F.}\ \bibnamefont {Herbut}},\ }\bibfield
   {title} {\bibinfo {title} {{Emergent Lorentz symmetry near fermionic quantum
  critical points in two and three dimensions}},\ }\href
  {https://doi.org/10.1007/JHEP04(2016)018} {\bibfield  {journal} {\bibinfo
  {journal} {JHEP}\ }\textbf {\bibinfo {volume} {04}},\ \bibinfo {pages}
  {018}},\ \Eprint {https://arxiv.org/abs/1510.07650} {arXiv:1510.07650
  [hep-th]} \BibitemShut {NoStop}%
\bibitem [{\citenamefont {Pozo}\ \emph {et~al.}(2018)\citenamefont {Pozo},
  \citenamefont {Ferreiros},\ and\ \citenamefont {Vozmediano}}]{Pozo:2018yzs}%
  \BibitemOpen
  \bibfield  {author} {\bibinfo {author} {\bibfnamefont {O.}~\bibnamefont
  {Pozo}}, \bibinfo {author} {\bibfnamefont {Y.}~\bibnamefont {Ferreiros}},\
  and\ \bibinfo {author} {\bibfnamefont {M.~A.~H.}\ \bibnamefont
  {Vozmediano}},\ }\bibfield  {title} {\bibinfo {title} {{Anisotropic fixed
  points in Dirac and Weyl semimetals}},\ }\href
  {https://doi.org/10.1103/PhysRevB.98.115122} {\bibfield  {journal} {\bibinfo
  {journal} {Phys. Rev.}\ }\textbf {\bibinfo {volume} {B98}},\ \bibinfo {pages}
  {115122} (\bibinfo {year} {2018})},\ \Eprint
  {https://arxiv.org/abs/1802.02632} {arXiv:1802.02632 [cond-mat.str-el]}
  \BibitemShut {NoStop}%
\bibitem [{\citenamefont {Chernodub}\ and\ \citenamefont
  {Vozmediano}(2019)}]{Chernodub:2019blw}%
  \BibitemOpen
  \bibfield  {author} {\bibinfo {author} {\bibfnamefont {M.~N.}\ \bibnamefont
  {Chernodub}}\ and\ \bibinfo {author} {\bibfnamefont {M.~A.~H.}\ \bibnamefont
  {Vozmediano}},\ }\bibfield  {title} {\bibinfo {title} {{Direct measurement of
  a beta function and an indirect check of the Schwinger effect near the
  boundary in Dirac-Weyl semimetals}},\ }\href
  {https://doi.org/10.1103/PhysRevResearch.1.032002} {\bibfield  {journal}
  {\bibinfo  {journal} {Phys. Rev. Research.}\ }\textbf {\bibinfo {volume}
  {1}},\ \bibinfo {pages} {032002} (\bibinfo {year} {2019})},\ \Eprint
  {https://arxiv.org/abs/1902.02694} {arXiv:1902.02694 [cond-mat.str-el]}
  \BibitemShut {NoStop}%
\bibitem [{\citenamefont {Landau}\ \emph {et~al.}(1984)\citenamefont {Landau},
  \citenamefont {Lifshitz},\ and\ \citenamefont {Pitaevskii}}]{Landau8}%
  \BibitemOpen
  \bibfield  {author} {\bibinfo {author} {\bibfnamefont {L.~D.}\ \bibnamefont
  {Landau}}, \bibinfo {author} {\bibfnamefont {E.~M.}\ \bibnamefont
  {Lifshitz}},\ and\ \bibinfo {author} {\bibfnamefont {L.~P.}\ \bibnamefont
  {Pitaevskii}},\ }\href@noop {} {\emph {\bibinfo {title} {Electrodynamics of
  continuous media}}}\ (\bibinfo  {publisher} {Pergamon, Oxford},\ \bibinfo
  {year} {1984})\BibitemShut {NoStop}%
\bibitem [{\citenamefont {Khusnutdinov}\ and\ \citenamefont
  {Emelianova}(2019)}]{Khusnutdinov:2018ley}%
  \BibitemOpen
  \bibfield  {author} {\bibinfo {author} {\bibfnamefont {N.}~\bibnamefont
  {Khusnutdinov}}\ and\ \bibinfo {author} {\bibfnamefont {N.}~\bibnamefont
  {Emelianova}},\ }\bibfield  {title} {\bibinfo {title} {{Low-temperature
  expansion of the Casimir–Polder free energy for an atom interacting with a
  conductive plane}},\ }\href {https://doi.org/10.1142/S0217751X19500088}
  {\bibfield  {journal} {\bibinfo  {journal} {Int. J. Mod. Phys.}\ }\textbf
  {\bibinfo {volume} {A34}},\ \bibinfo {pages} {1950008} (\bibinfo {year}
  {2019})},\ \Eprint {https://arxiv.org/abs/1808.08261} {arXiv:1808.08261
  [cond-mat.mes-hall]} \BibitemShut {NoStop}%
\bibitem [{\citenamefont {Kurkov}\ and\ \citenamefont
  {Vassilevich}(2020)}]{Kurkov:2020jet}%
  \BibitemOpen
  \bibfield  {author} {\bibinfo {author} {\bibfnamefont {M.}~\bibnamefont
  {Kurkov}}\ and\ \bibinfo {author} {\bibfnamefont {D.}~\bibnamefont
  {Vassilevich}},\ }\bibfield  {title} {\bibinfo {title} {{How many surface
  modes does one see on the boundary of a Dirac material?}},\ }\href
  {https://doi.org/10.1103/PhysRevLett.124.176802} {\bibfield  {journal}
  {\bibinfo  {journal} {Phys. Rev. Lett.}\ }\textbf {\bibinfo {volume} {124}},\
  \bibinfo {pages} {176802} (\bibinfo {year} {2020})},\ \Eprint
  {https://arxiv.org/abs/2002.06721} {arXiv:2002.06721 [hep-th]} \BibitemShut
  {NoStop}%
\bibitem [{\citenamefont {Reuter}\ and\ \citenamefont
  {Sondheimer}(1948)}]{Reuter:1948}%
  \BibitemOpen
  \bibfield  {author} {\bibinfo {author} {\bibfnamefont {G.~E.~H.}\
  \bibnamefont {Reuter}}\ and\ \bibinfo {author} {\bibfnamefont {E.~H.}\
  \bibnamefont {Sondheimer}},\ }\bibfield  {title} {\bibinfo {title} {{The
  theory of anomalous skin effect in metals}},\ }\href
  {https://doi.org/10.1098/rspa.1948.0123} {\bibfield  {journal} {\bibinfo
  {journal} {Proc. Royal Soc. London, Ser. A}\ }\textbf {\bibinfo {volume}
  {195}},\ \bibinfo {pages} {336} (\bibinfo {year} {1948})}\BibitemShut
  {NoStop}%
\bibitem [{\citenamefont {Silin}\ and\ \citenamefont
  {Fetisov}(1962)}]{Silin:1962}%
  \BibitemOpen
  \bibfield  {author} {\bibinfo {author} {\bibfnamefont {V.~P.}\ \bibnamefont
  {Silin}}\ and\ \bibinfo {author} {\bibfnamefont {E.~P.}\ \bibnamefont
  {Fetisov}},\ }\bibfield  {title} {\bibinfo {title} {Electromagnetic
  properties of a relativistic plasma},\ }\href@noop {} {\bibfield  {journal}
  {\bibinfo  {journal} {Sov. Phys. JETP}\ }\textbf {\bibinfo {volume} {14}},\
  \bibinfo {pages} {115} (\bibinfo {year} {1962})}\BibitemShut {NoStop}%
\bibitem [{\citenamefont {Kliewer}\ and\ \citenamefont
  {Fuchs}(1968)}]{Kliewer:1968zz}%
  \BibitemOpen
  \bibfield  {author} {\bibinfo {author} {\bibfnamefont {K.~L.}\ \bibnamefont
  {Kliewer}}\ and\ \bibinfo {author} {\bibfnamefont {R.}~\bibnamefont
  {Fuchs}},\ }\bibfield  {title} {\bibinfo {title} {{Anomalous Skin Effect for
  Specular Electron Scattering and Optical Experiments at Non-Normal Angles of
  Incidence}},\ }\href {https://doi.org/10.1103/PhysRev.172.607} {\bibfield
  {journal} {\bibinfo  {journal} {Phys. Rev.}\ }\textbf {\bibinfo {volume}
  {172}},\ \bibinfo {pages} {607} (\bibinfo {year} {1968})}\BibitemShut
  {NoStop}%
\bibitem [{\citenamefont {Esquivel}\ and\ \citenamefont
  {Svetovoy}(2004)}]{Esquivel:2004zz}%
  \BibitemOpen
  \bibfield  {author} {\bibinfo {author} {\bibfnamefont {R.}~\bibnamefont
  {Esquivel}}\ and\ \bibinfo {author} {\bibfnamefont {V.~B.}\ \bibnamefont
  {Svetovoy}},\ }\bibfield  {title} {\bibinfo {title} {{Correction to the
  Casimir force due to the anomalous skin effect}},\ }\href
  {https://doi.org/10.1103/PhysRevA.69.062102} {\bibfield  {journal} {\bibinfo
  {journal} {Phys. Rev.}\ }\textbf {\bibinfo {volume} {A69}},\ \bibinfo {pages}
  {062102} (\bibinfo {year} {2004})},\ \Eprint
  {https://arxiv.org/abs/quant-ph/0404073} {arXiv:quant-ph/0404073 [quant-ph]}
  \BibitemShut {NoStop}%
\bibitem [{\citenamefont {Bordag}\ \emph {et~al.}(2016)\citenamefont {Bordag},
  \citenamefont {Fialkovskiy},\ and\ \citenamefont
  {Vassilevich}}]{Bordag:2015zda}%
  \BibitemOpen
  \bibfield  {author} {\bibinfo {author} {\bibfnamefont {M.}~\bibnamefont
  {Bordag}}, \bibinfo {author} {\bibfnamefont {I.}~\bibnamefont
  {Fialkovskiy}},\ and\ \bibinfo {author} {\bibfnamefont {D.}~\bibnamefont
  {Vassilevich}},\ }\bibfield  {title} {\bibinfo {title} {{Enhanced Casimir
  effect for doped graphene}},\ }\href
  {https://doi.org/10.1103/PhysRevB.93.075414} {\bibfield  {journal} {\bibinfo
  {journal} {Phys. Rev.}\ }\textbf {\bibinfo {volume} {B 93}},\ \bibinfo
  {pages} {075414} (\bibinfo {year} {2016})},\ \Eprint
  {https://arxiv.org/abs/1507.08693} {arXiv:1507.08693 [cond-mat.mes-hall]}
  \BibitemShut {NoStop}%
\bibitem [{\citenamefont {Shuryak}(1980)}]{Shuryak:1980tp}%
  \BibitemOpen
  \bibfield  {author} {\bibinfo {author} {\bibfnamefont {E.~V.}\ \bibnamefont
  {Shuryak}},\ }\bibfield  {title} {\bibinfo {title} {{Quantum Chromodynamics
  and the Theory of Superdense Matter}},\ }\href
  {https://doi.org/10.1016/0370-1573(80)90105-2} {\bibfield  {journal}
  {\bibinfo  {journal} {Phys. Rept.}\ }\textbf {\bibinfo {volume} {61}},\
  \bibinfo {pages} {71} (\bibinfo {year} {1980})}\BibitemShut {NoStop}%
\bibitem [{\citenamefont {Bordag}\ \emph {et~al.}(2015)\citenamefont {Bordag},
  \citenamefont {Klimchitskaya}, \citenamefont {Mostepanenko},\ and\
  \citenamefont {Petrov}}]{Bordag:2015rqa}%
  \BibitemOpen
  \bibfield  {author} {\bibinfo {author} {\bibfnamefont {M.}~\bibnamefont
  {Bordag}}, \bibinfo {author} {\bibfnamefont {G.~L.}\ \bibnamefont
  {Klimchitskaya}}, \bibinfo {author} {\bibfnamefont {V.~M.}\ \bibnamefont
  {Mostepanenko}},\ and\ \bibinfo {author} {\bibfnamefont {V.~M.}\ \bibnamefont
  {Petrov}},\ }\bibfield  {title} {\bibinfo {title} {{Quantum field theoretical
  description for the reflectivity of graphene}},\ }\href
  {https://doi.org/10.1103/PhysRevD.93.089907, 10.1103/PhysRevD.91.045037,
  10.1103/PhysRevD.91.040537} {\bibfield  {journal} {\bibinfo  {journal} {Phys.
  Rev.}\ }\textbf {\bibinfo {volume} {D91}},\ \bibinfo {pages} {045037}
  (\bibinfo {year} {2015})},\ \bibinfo {note} {[Erratum: Phys.
  Rev.D93,no.8,089907(2016)]},\ \Eprint {https://arxiv.org/abs/1501.07715}
  {arXiv:1501.07715 [cond-mat.mes-hall]} \BibitemShut {NoStop}%
\bibitem [{\citenamefont {Bordag}\ and\ \citenamefont
  {Pirozhenko}(2015)}]{Bordag:2015gla}%
  \BibitemOpen
  \bibfield  {author} {\bibinfo {author} {\bibfnamefont {M.}~\bibnamefont
  {Bordag}}\ and\ \bibinfo {author} {\bibfnamefont {I.~G.}\ \bibnamefont
  {Pirozhenko}},\ }\bibfield  {title} {\bibinfo {title} {{Surface plasmons for
  doped graphene}},\ }\href {https://doi.org/10.1103/PhysRevD.91.085038}
  {\bibfield  {journal} {\bibinfo  {journal} {Phys. Rev. D}\ }\textbf {\bibinfo
  {volume} {91}},\ \bibinfo {pages} {085038} (\bibinfo {year} {2015})},\
  \Eprint {https://arxiv.org/abs/1502.00421} {arXiv:1502.00421
  [cond-mat.mes-hall]} \BibitemShut {NoStop}%
\bibitem [{\citenamefont {Falomir}\ \emph {et~al.}(2020)\citenamefont
  {Falomir}, \citenamefont {Mu\~noz}, \citenamefont {Loewe},\ and\
  \citenamefont {Zamora}}]{Falomir:2019yfk}%
  \BibitemOpen
  \bibfield  {author} {\bibinfo {author} {\bibfnamefont {H.}~\bibnamefont
  {Falomir}}, \bibinfo {author} {\bibfnamefont {E.}~\bibnamefont {Mu\~noz}},
  \bibinfo {author} {\bibfnamefont {M.}~\bibnamefont {Loewe}},\ and\ \bibinfo
  {author} {\bibfnamefont {R.}~\bibnamefont {Zamora}},\ }\bibfield  {title}
  {\bibinfo {title} {{Optical Conductivity in an effective model for Graphene:
  Finite temperature corrections}},\ }\href
  {https://doi.org/10.1088/1751-8121/ab57cb} {\bibfield  {journal} {\bibinfo
  {journal} {J. Phys. A}\ }\textbf {\bibinfo {volume} {53}},\ \bibinfo {pages}
  {015401} (\bibinfo {year} {2020})},\ \Eprint
  {https://arxiv.org/abs/1907.02017} {arXiv:1907.02017 [cond-mat.mes-hall]}
  \BibitemShut {NoStop}%
\end{thebibliography}

\end{document}